\documentclass[manuscript,nonacm]{acmart}
\AtBeginDocument{%
  }

\setcopyright{none}

\acmISBN{978-1-4503-XXXX-X/2018/06}

\usepackage{makecell}
\usepackage{multirow} 
\usepackage{placeins}
\usepackage[most]{tcolorbox}

\newtcolorbox{promptbox}[1]{
    breakable,
    colback=black!2,
    colframe=black!25,
    boxrule=0.35pt,
    arc=1pt,
    left=3pt,
    right=3pt,
    top=3pt,
    bottom=3pt,
    title={#1},
    fonttitle=\bfseries\footnotesize,
    fontupper=\scriptsize,
    before upper={
        \setlength{\parindent}{0pt}
        \setlength{\parskip}{2pt}
    },
    before skip=4pt,
    after skip=6pt
}
\begin{document}

\title{CHORAL: A Multi-Agent Scaffold for Longitudinal Prognostic Reasoning from Patient-Generated Health Data}
\renewcommand{\shorttitle}{CHORAL}

\author{Youngji Koh}
\affiliation{%
  \institution{KAIST}
  \city{Daejeon}
  \country{South Korea}
}
\orcid{0000-0001-8405-4919}
\email{youngji@kaist.ac.kr}

\author{Esther Kang}
\affiliation{%
  \institution{KAIST}
  \city{Daejeon}
  \country{South Korea}}
\email{esther.kang@kaist.ac.kr}
\orcid{0009-0005-5992-072X}

\author{Gyuna Kim}
\affiliation{%
  \institution{KAIST}
  \city{Daejeon}
  \country{South Korea}}
\email{gyuna.kim@kaist.ac.kr}

\author{Seungwan Jin}
\affiliation{%
  \institution{KAIST}
  \city{Daejeon}
  \country{South Korea}}
\email{seungwan@kaist.ac.kr}

\author{Hyunsoo Lee}
\affiliation{%
  \institution{KAIST}
  \city{Daejeon}
  \country{South Korea}}
\email{hslee90@kaist.ac.kr}

\author{Kyungsik Han}
\affiliation{%
  \institution{Hanyang University}
  \city{Seoul}
  \country{South Korea}}
\email{kyungsikhan@hanyang.ac.kr}

\author{Uichin Lee}
\authornote{Corresponding author}
\affiliation{%
  \institution{KAIST}
  \city{Daejeon}
  \country{South Korea}
}
\orcid{0000-0002-1888-1569}
\email{uclee@kaist.ac.kr}

\renewcommand{\shortauthors}{Koh et al.}

\begin{abstract}
Longitudinal patient-generated health data can inform prognostic reasoning about how patients' conditions may evolve between follow-up visits. Existing approaches summarize trajectories or predict symptom states, but provide limited support for examining how heterogeneous evidence informs prognosis. We present CHORAL, a multi-agent scaffold that forms domain-specific appraisals before cross-domain synthesis and review, preserving evidence and revisions. Informed by a formative study with 12 psychiatrists, we evaluated CHORAL quantitatively and with 22 clinicians. CHORAL achieved balanced accuracy comparable to persistence (0.673 vs. 0.672), while detecting 40\% of LOW-to-HIGH deterioration cases missed by persistence. Versus a single-agent baseline, it improved balanced accuracy from 0.639 to 0.673 and evidence coverage from 70.1\% to 89.2\%. Clinicians rated CHORAL significantly higher for evidence appraisal, alternative interpretations, grounds for challenge, and distinguishing what to rely on from what required verification. These findings show how multi-agent deliberation can support review by exposing alternative interpretations and grounds for verification.
\end{abstract}




\maketitle
\section{Introduction}
Mental health outpatient care often involves repeated follow-up visits, during which clinicians assess how a patient's condition has changed since the previous encounter and prepare for subsequent care \cite{wagner2001improving}. To do so, clinicians consider not only the patient's current state (\textit{diagnosis}) but also whether the recent clinical course is worsening, improving, persisting, or fluctuating, and anticipate possible future clinical courses (\textit{prognosis}). Such \textit{prognostic reasoning} is a forward-oriented inference that connects diagnostic reasoning about the patient's current state with therapeutic reasoning about subsequent intervention \cite{higgs2024clinical,nurcombe1986clinical}. 

However, outpatient follow-up visits are typically separated by multiple weeks, making it difficult to continuously observe patients' everyday behavioral trajectories between encounters \cite{hong2021implementing}. Patient-generated health data (PGHD) can help fill this gap by longitudinally capturing behaviors outside the clinic, such as physical activity, sleep, mobility, and smartphone use \cite{mouchabac2021improving,insel2017digital,torous2020digital}. In mental health care, these behavioral changes can provide trajectory-level evidence for examining potential future concerns related to depression, anxiety, stress, and sleep problems \cite{fortney2017tipping,trivedi2006evaluation}. Yet, using longitudinal PGHD for prognostic reasoning remains challenging.

First, constructing a meaningful longitudinal trajectory from heterogeneous PGHD is difficult. PGHD consists of multiple behavioral signals with different scales and temporal patterns, and clinicians must distinguish transient fluctuations from persistent changes. This requires considering the magnitude, persistence, and direction of multiple signals relative to each patient's own baseline \cite{fisher2015toward}. Second, the meaning of an observed behavioral trajectory may differ across areas relevant to mental health, such as depression, anxiety, stress, and sleep problems. The same behavioral change can be relevant to more than one of these domains, making it difficult to understand a patient's condition from a single domain-specific perspective \cite{choi2024digital}. Prognostic reasoning therefore requires examining the same longitudinal evidence across multiple domains and considering how the resulting domain-specific prognostic appraisals relate to one another, rather than reducing the trajectory to a single interpretation. Third, PGHD alone cannot fully explain the clinical context of behavioral change. Assessing the significance of an observed behavioral change requires considering its relevance within different domains as well as its potential relevance to everyday functioning \cite{american2022diagnostic,ustun2010measuring}. However, the contextual information needed to fully interpret such changes is not necessarily contained in PGHD. PGHD may show \textit{what} changed, but not why the change occurred or how the patient experienced it. Systems should therefore distinguish what can be interpreted from the available evidence from information gaps that require additional clinical context during follow-up.

Prior work in HCI and mental health sensing has partly addressed these challenges through PGHD summarization, visualization, and prediction. Summarization and visualization organize longitudinal behavioral signals to help clinicians understand \textit{what happened} \cite{zou2026mind,xu2026lens}, but provide limited support for examining how observed changes may raise different prognostic concerns across domains. Prediction approaches estimate future outcomes or risks from PGHD \cite{englhardt2024classification,kim2024health}, but typically center on the final prediction rather than how longitudinal evidence is interpreted across domains. Clinical multi-agent systems have also distributed reasoning across specialized roles or stages \cite{kim2024mdagents,tang2024medagents,wang2025colacare}, but provide limited support for preserving separate appraisals of the same longitudinal evidence before cross-domain integration. Thus, a gap remains in scaffolding the \textit{multi-domain prognostic reasoning process itself} over longitudinal PGHD.

To scaffold PGHD-based prognostic reasoning, clinicians should be able to examine not only a system's final output but also the evidence and reasoning underlying it. When a system's interpretation differs from their own assessment, clinicians should be able to inspect its rationale and, when necessary, question or contest the judgment \cite{alfrink2023contestable,lyons2021conceptualising}. Intermediate reasoning should therefore remain inspectable, including the evidence supporting domain-specific prognostic appraisals and how these appraisals are integrated across domains.

Based on this motivation, we developed \textbf{CHORAL} (\textbf{\underline{C}}ontext-aware \textbf{\underline{H}}ealth-data \textbf{\underline{O}}rchestration of \textbf{\underline{R}}easoning \textbf{\underline{A}}gents for \textbf{\underline{L}}ongitudinal care), a multi-agent reasoning scaffold for longitudinal prognostic reasoning in mental health follow-up care. We conducted a formative study with 12 psychiatrists to identify key reasoning functions involved in psychiatric follow-up and translated these functions into the system design. CHORAL constructs longitudinal trajectories, forms separate prognostic appraisals for depression, anxiety, stress, and sleep, identifies information gaps, and integrates the resulting appraisals through cross-domain synthesis. Each domain appraisal is formed before access to the other domain appraisals, allowing differences in interpretation to emerge before synthesis. Subsequent review and revision keep supporting evidence, disagreements, and changes explicit and inspectable.

We investigate the following research questions:

\begin{itemize}
\item \textbf{RQ1.} How do CHORAL's architectural components contribute to prognostic performance, evidence use, and citation validity over longitudinal PGHD?
\item \textbf{RQ2.} How do clinicians perceive CHORAL's usefulness for interpreting longitudinal PGHD and preparing follow-up plans?
\item \textbf{RQ3.} How does CHORAL support clinicians in contesting prognostic judgments and deciding what to rely on or verify?
\end{itemize}

To address these questions, we conducted two complementary evaluations. We first evaluated CHORAL on a longitudinal mental health dataset against majority, persistence, and single-agent baselines, together with component ablations (RQ1). We then conducted a within-subject study with 22 psychiatry clinicians comparing CHORAL with the single-agent baseline to examine clinical usefulness, follow-up support, reasoning review, and reliance support (RQ2--RQ3).

Our findings show that CHORAL achieved balanced accuracy comparable to the persistence baseline (.673 vs. .672), while detecting 40\% of LOW-to-HIGH deterioration cases that persistence did not detect. Compared with the single-agent baseline, CHORAL also achieved higher balanced accuracy (.673 vs. .639) and broader evidence channel coverage (89.2\% vs. 70.1\%). In the clinician study, CHORAL was particularly valued for supporting evidence appraisal, surfacing alternative interpretations, and providing concrete grounds for challenging system judgments. It also helped clinicians distinguish which aspects of the system output could be relied on and which required further verification. Cross-domain interpretation was rated more highly under CHORAL, although this difference did not remain significant after Benjamini--Hochberg correction. In contrast, these benefits did not clearly extend to follow-up prioritization or workflow integration. Building on these findings, we discuss design implications for clinician-centered AI systems that support inspectable and contestable longitudinal prognostic reasoning in mental health follow-up care.

In summary, our contributions are as follows:

\begin{itemize}
\item Through a formative study with 12 psychiatrists, we derive design requirements for longitudinal, multi-domain prognostic reasoning and instantiate them in CHORAL, which preserves separate domain appraisals, distinguishes retrievable from externally unresolved information gaps, and keeps synthesis and revision traceable for clinician review.
\item Through system-level and clinician evaluations, we show that CHORAL supports broader evidence use and helps clinicians identify alternative interpretations, locate grounds for contestation, and distinguish what requires further verification.
\item We derive design implications for surfacing consequential deliberation while separating reasoning depth from presentation depth.
\end{itemize}
\section{Background and Related Work}
In this section, we review existing literature on longitudinal prognostic reasoning, AI support for clinical reasoning, and multi-agent systems for structuring clinical judgment.

\subsection{Prognostic Reasoning in Mental Health Follow-up}

Prognostic reasoning concerns how a patient's condition may evolve and how that anticipated course should inform subsequent care \cite{higgs2024clinical,nurcombe1986clinical}. In follow-up care, such judgments are longitudinal and provisional: clinicians compare the current state with the patient's prior course, assess the direction and persistence of change, and revise earlier interpretations as new evidence becomes available \cite{heneghan2009diagnostic,ilgen2016s}.

Longitudinal patient-generated health data (PGHD) can provide additional evidence for this process by capturing behavioral changes between clinical encounters. Prior sensing research has linked mobility and other behavioral signals to depressive symptoms \cite{saeb2015mobile,canzian2015trajectories}, while recent digital phenotyping studies have shown substantial variation in behavioral and physiological patterns across individuals and over time \cite{lampe2026interindividual,zhan2025digital,douglas2026digital}. Together, this work suggests that PGHD can provide useful evidence about changes in a patient's behavioral state across time.

However, interpreting the clinical meaning of these behavioral changes remains difficult in mental health. Psychiatric symptoms frequently overlap and co-occur, making it difficult to interpret an observed change from a single symptom perspective \cite{borsboom2017network,insel2010research}. The same behavioral manifestation may support multiple symptom-related or non-clinical interpretations. Assessing the clinical significance of such changes also requires considering their relevance to everyday functioning \cite{american2022diagnostic,ustun2010measuring}. Subjective experience and situational context are also important for distinguishing among possible interpretations, but such information is not necessarily captured in passively sensed or structured data.

Accordingly, using PGHD for prognosis requires interpreting change in relation to the patient's prior course, considering multiple plausible explanations, identifying missing contextual information, and updating judgments as new evidence becomes available. Carrying out these steps across heterogeneous longitudinal data can impose substantial interpretive burden within time-constrained care \cite{tonekaboni2019clinicians,sivaraman2026intelligent}. This creates a need for tools that support these intermediate reasoning steps rather than only presenting longitudinal data or a final prediction.

\subsection{AI Scaffolding for Prognostic Reasoning}

Prior AI and HCI systems have helped users interpret longitudinal health data by organizing multimodal sensing streams into patterns, narratives, and visualizations \cite{choube2025gloss,zou2026mind,xu2026lens}. These systems make behavioral trajectories more accessible, but primarily focus on understanding observed or current states.

Complementary work has used longitudinal sensing and health-record data to estimate future mental-health and clinical states \cite{stamatis2024differential,balliu2024personalized,zhu2026augmenting,kim2024health}. These approaches extend longitudinal analysis from understanding observed states toward anticipating future outcomes. However, even when predictions are accompanied by supporting evidence, clinicians must still determine how that evidence should be interpreted, what alternative interpretations remain plausible, and what additional information may be needed before forming a clinical judgment.

Recent HCI research has increasingly examined how AI can support the human reasoning process rather than only provide a final recommendation. Sivaraman et al.\ conceptualized AI-provided information as \emph{reasoning cues}, while Zhang et al.\ contrasted recommendation-centric AI with continuous support throughout decision making \cite{sivaraman2026intelligent,zhang2024beyond}. This distinction is particularly important given evidence that simply providing physicians with access to a capable LLM does not necessarily improve their diagnostic reasoning compared with conventional resources \cite{goh2024large}. Together, these findings motivate AI that supports how judgments are formed rather than simply delivering answers.

Clinical reasoning support also raises the question of whether clinicians can inspect and question the basis of AI-supported judgments. Explainable AI research has emphasized the importance of making algorithmic decisions understandable \cite{miller2019explanation}, while contestable AI has argued that users should be able to critically engage with and challenge algorithmic judgments \cite{lyons2021conceptualising,alfrink2023contestable}. A recent psychiatric AI system, Heart2Mind, operationalizes this direction by combining explanations with clinician--AI dialogue to support examination and contestation of model predictions \cite{nguyen2026heart2mind}. However, this work largely focuses on scrutinizing an AI output and its basis, with less attention to how a judgment develops through intermediate reasoning over longitudinal evidence.

\subsection{Multi-Agent AI for Clinical Reasoning}

Recent clinical AI research has increasingly used multi-agent systems to structure complex reasoning across specialized roles and perspectives \cite{tang2024medagents,kim2024mdagents,chen2025mdteamgpt}. Early approaches largely emphasized specialization, task decomposition, and aggregation across expert roles \cite{tang2024medagents,wang2024beyond,kim2024mdagents,wang2026medagent}. Later systems explicitly structure collaboration and synthesis through adaptive configurations, lead-agent orchestration, or meta-level coordination \cite{kim2024mdagents,chen2025mdteamgpt,li2026med,wang2025colacare}. These studies illustrate a shift from assigning specialized roles toward coordinating how distributed judgments are integrated.

In mental health, multi-agent systems have mainly distributed distinct clinical functions or stages, including interviewing, screening, diagnostic assessment, and patient communication \cite{bi2025magi,wu2026wisemind,kamaleddin2026multi,greene2025ai,tang2025layered}. In contrast, multi-domain prognosis requires the same longitudinal trajectory to be appraised from multiple symptom-domain perspectives before integration, because the same behavioral pattern may have different implications for depression, anxiety, stress, and sleep \cite{borsboom2017network,insel2010research,choi2024digital}. Prognostic reasoning therefore requires preserving domain-specific appraisals before integration while allowing provisional judgments to be revisited as new evidence becomes available \cite{heneghan2009diagnostic,ilgen2016s}.

The challenge is therefore not merely to generate multiple appraisals, but to structure how they are grounded in evidence, synthesized across domains, and reconsidered through review. Prior systems have supported multi-LLM sensemaking of passive sensing data (e.g., GLOSS~\cite{choube2025gloss}), clinician-facing narrative interpretation of multimodal mental-health data (e.g., MIND~\cite{zou2026mind}), and multi-agent collaboration over clinical records (e.g., ColaCare~\cite{wang2025colacare}). CHORAL instead focuses on preserving how prognostic interpretations evolve over the same longitudinal evidence: separate domain appraisals are retained before synthesis, information gaps are distinguished by whether they can be resolved from the record, and subsequent review or revision remains traceable to clinicians.
\section{System Design}
CHORAL was designed based on the clinical reasoning and system support needs identified through a formative study. This section first presents the formative study and the resulting design requirements, followed by the CHORAL architecture developed to operationalize them.

\subsection{Formative Study}
Our goal was to design a system that supports deliberative clinical reasoning with longitudinal patient information in mental health follow-up care. We therefore conducted a formative study to understand how clinicians interpret longitudinal changes and translate them into clinical judgments, and how a system could support this process.

\subsubsection{Study Design}
We conducted semi-structured interviews with 12 psychiatrists recruited from university hospitals and clinics across South Korea (4 female; age $M$=44.5, $SD$=6.43). Clinical experience was defined as years of practice after board certification: two participants had 0--5 years of experience, two had 6--10 years, and eight had more than 10 years. Participants were recruited through snowball sampling, beginning with a small group of psychiatrists and expanding through professional referrals, an approach commonly used to recruit hard-to-reach clinical populations in HCI research~\cite{jung2026toward}.

Each interview lasted approximately 60 minutes and was conducted online. The interviews focused on two broad topics: (1) how clinicians reason during mental health follow-up care, and (2) how a system could support reasoning with longitudinal patient information. Interviews were audio-recorded with participants' consent, and each participant received approximately USD 150 as compensation. The study was approved by the Institutional Review Board (IRB), and all participants provided written informed consent before participation. 

Two researchers independently coded the interview transcripts using thematic analysis. The researchers then organized and refined the emerging themes through affinity diagramming and discussion until consensus was reached. Through this analysis, we identified clinical reasoning practices in longitudinal follow-up care as well as clinicians' needs for system support.

\subsubsection{Findings and Design Requirements}
Our analysis identified four key reasoning functions in longitudinal follow-up care:
(1) \textit{tracing longitudinal change},
(2) \textit{appraising longitudinal change across multiple domains},
(3) \textit{identifying information gaps}, and
(4) \textit{integrating domain-specific appraisals for follow-up}.
Clinicians also emphasized that the system should support rather than replace clinical judgment, make the basis of system-generated judgments inspectable, and present information in a form that can be reviewed within limited follow-up time.
Based on these findings, we derived five design requirements, summarized with representative participant quotes in Table~\ref{tab:formative-findings}.

\begin{table*}[t]
\centering
\caption{Formative findings, representative participant quotes, and corresponding design requirements.}
\label{tab:formative-findings}

\begin{tabular}{
>{\raggedright\arraybackslash}p{0.27\textwidth}
>{\raggedright\arraybackslash}p{0.42\textwidth}
>{\raggedright\arraybackslash}p{0.25\textwidth}
}
\toprule
\textbf{Formative finding} & \textbf{Representative quotes} & \textbf{Design requirement} \\
\midrule

Clinicians examined changes relative to the patient's prior trajectory
&
\textbf{P3:} ``I first ask whether this is actually different from how this patient usually is...''\par
\textbf{P8:} ``I first look at the direction of change, whether the patient is worsening, improving, or only fluctuating briefly.''
&
\textbf{DR1.} Establish longitudinal trajectory before domain-specific appraisal
\\
\midrule

The same behavioral change was considered across multiple domains
&
\textbf{P4:} ``Most patients don't fit neatly into a single diagnosis...''\par
\textbf{P9:} ``When I see reduced activity, I do not immediately think depression... I usually look at several symptom areas before deciding what the change means.''
&
\textbf{DR2.} Appraise longitudinal change across multiple domains
\\
\midrule

Unclear interpretations prompted clinicians to seek missing information
&
\textbf{P6:} ``If something does not fit, I may look at the previous course or ask a family member...''\par
\textbf{P10:} ``Sometimes the data tells me that something changed, but not why... I would want to check what happened during that period before making a judgment.''
&
\textbf{DR3.} Identify information gaps and support targeted follow-up
\\
\midrule

Clinicians integrated domain-specific appraisals while comparing how they complemented or conflicted with one another
&
\textbf{P1:} ``I put those pieces together to understand the overall picture...''\par
\textbf{P8:} ``I look at how they fit together, where they conflict, and which concerns should take priority.''\par
\textbf{P7:}``I may think the depression is improving while the sleep problem is getting worse... one domain alone does not tell me the whole story.''
&
\textbf{DR4.} Integrate perspectives while preserving traceability
\\
\midrule

Limited visit time required prioritized information with details available on demand
&
\textbf{P1:} ``There simply is not time to read through lengthy text.''\par
\textbf{P11:} ``It would be more efficient if the summary came first...''\par
\textbf{P12:} ``If the system gives me everything at once, I probably will not read it during a follow-up visit... I would rather see the main concerns first...''
&
\textbf{DR5.} Prioritize information for time-constrained follow-up
\\

\bottomrule
\end{tabular}
\end{table*}

\paragraph{DR1: Establish longitudinal trajectory before domain-specific appraisal.}
Clinicians first assessed change relative to the patient's prior state, focusing on its direction and persistence before domain-specific interpretation. Some related this sequence to SOAP-style documentation, in which observations are organized before clinical assessment~\cite{weed1968medical}. Accordingly, the system should first establish a patient-level longitudinal trajectory before domain-specific appraisal.

\paragraph{DR2: Appraise longitudinal change across multiple domains.}
Clinicians considered the same behavioral change from multiple domain-specific perspectives rather than assigning it directly to a single interpretation. Accordingly, the system should preserve separate domain-specific appraisals before integrating them at the case level.

\paragraph{DR3: Identify information gaps and support targeted follow-up.}
When observed changes were difficult to interpret, clinicians sought additional context from prior history, the patient, or other sources. The system should therefore retrieve relevant longitudinal evidence when available and surface unresolved information needs for subsequent clinical follow-up.

\paragraph{DR4: Integrate domain-specific appraisals while preserving traceability.}
Clinicians formed an overall understanding by comparing how domain-specific appraisals complemented or conflicted with one another while retaining their supporting evidence. The system should therefore integrate these appraisals while preserving the evidence and reasoning underlying each interpretation.

\paragraph{DR5: Prioritize information for time-constrained follow-up.}
Clinicians emphasized the need to prioritize recent changes, major concerns, and unresolved issues within limited follow-up time. The system should therefore present key information first while keeping supporting evidence and reasoning available on demand.

These design requirements informed the reasoning mechanisms and interface design of CHORAL described in the following section.

\subsection{CHORAL System}

\begin{figure*}[t]
    \centering
    \includegraphics[width=\textwidth]{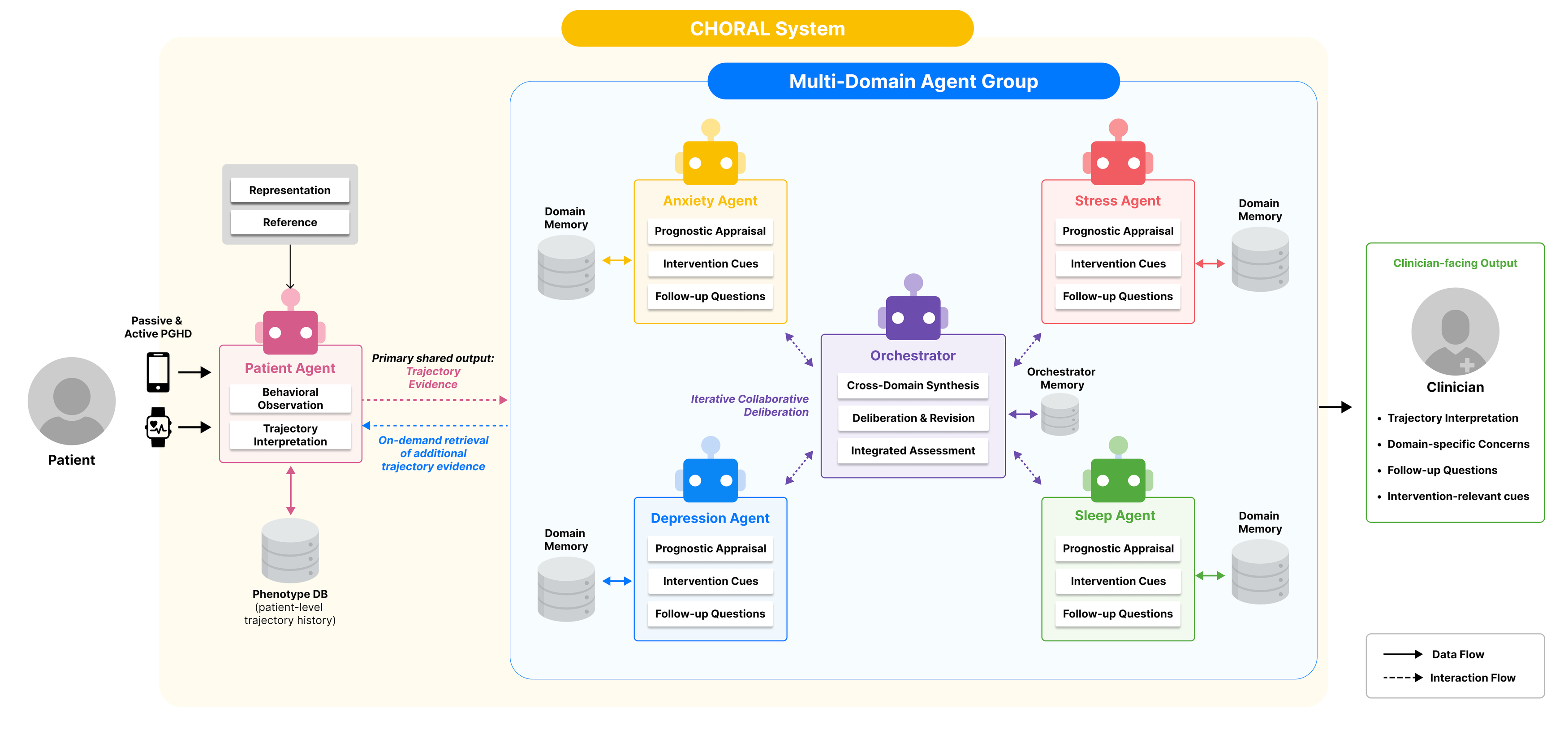}
    \caption{Overview of the CHORAL system architecture and reasoning flow.}
    \label{fig:system-overview}
\end{figure*}

Based on the five design requirements derived from our formative study, we developed CHORAL, a multi-agent system for supporting clinical reasoning over longitudinal PGHD. Figure~\ref{fig:system-overview} provides an overview of the system architecture and reasoning flow. Rather than producing a single end-to-end judgment, CHORAL organizes the reasoning process into several stages. The Patient Agent first constructs a shared longitudinal trajectory from the patient's PGHD. Domain Agents then examine the same trajectory independently from the perspectives of depression, anxiety, stress, and sleep. When additional information is needed, CHORAL either retrieves relevant information from the available longitudinal history or retains the unresolved information need for subsequent clinical follow-up. The Orchestrator integrates the domain-specific prognostic appraisals at the case level while preserving their evidence and differences across perspectives. Finally, the Clinical Dashboard presents the main findings first while keeping detailed evidence and reasoning available for further inspection. Implementation details and prompt templates are provided in Appendix~\ref{appendix-prompts}.

\subsubsection{Patient Agent}

To support DR1, the patient agent constructs a shared longitudinal trajectory from longitudinal PGHD. In the formative study, clinicians emphasized examining how a patient's behavior had changed relative to their usual state, whether the change was transient or persistent, and whether it could be relevant to everyday functioning. Rather than directly interpreting raw PGHD in relation to a specific domain, the patient agent first structures behavioral changes and their functioning-relevant context as patient-level evidence that can subsequently be interpreted from multiple domain-specific perspectives.

Constructing a patient-specific longitudinal trajectory requires considering what has changed and how, how different the current state is from the patient's usual pattern, and how the current change relates to the patient's prior course. We organize this process around three mechanisms: \textit{Representation}, \textit{Reference}, and \textit{Context}.

\textbf{Representation: Explicit behavioral trajectory representation.}
Longitudinal PGHD consists of heterogeneous signals such as activity, mobility, sleep, and phone use. Raw values alone do not provide a consistent description of the patient's behavioral state or how it changes over time. Prior work on sensing representations has likewise shown the importance of representing temporal patterns and changes in addition to raw sensor values~\cite{zhang2025sensorlm}. We therefore structure PGHD in terms of statistical characteristics, changes over time, and time-of-day patterns.

For each PGHD stream, statistical features summarize the overall state and variability during the observation period using the mean, standard deviation, minimum and maximum, median, and first and last observed values. Structural features describe the direction and magnitude of change using slope, trend direction, delta, percentage change, and day-to-day variability. Temporal features capture time-of-day behavioral patterns using four 6-hour periods, following prior sensing studies: night (00:00--06:00), morning (06:00--12:00), afternoon (12:00--18:00), and evening (18:00--24:00)~\cite{xia2025longitudinal}.

To represent functioning-relevant context, these behavioral features are also organized using a conceptual mapping between available sensing features and functioning-relevant domains informed by the WHO Disability Assessment Schedule (WHODAS) \cite{ustun2010measuring}. The full feature-to-domain mapping is provided in Appendix~\ref{appendix-mapping}. This mapping supports interpretation of potentially functioning-relevant behavioral changes rather than inferring WHODAS scores or establishing functional impairment.

\textbf{Reference: Baseline-relative trajectory construction.}
The same behavioral value can have different meanings across individuals because each patient has a different usual pattern. Current behavior therefore needs to be interpreted in relation to within-person change rather than by absolute level alone~\cite{fisher2015toward}. Clinicians in the formative study similarly described comparing the current state with both recent behavior and a more stable historical pattern when assessing whether a meaningful change had occurred.

For each reasoning round, the current state was summarized over a 7-day window ($t-6$ to $t$). For both passive-sensing and EMA data, the recent baseline was computed from the preceding 28-day period excluding the current window ($t-34$ to $t-7$). For passive-sensing features, the stable baseline was computed from all available sensing history preceding the current window, excluding sensing periods associated with EMA scores meeting the baseline-exclusion criteria (stress $\geq 2$, PHQ-2 $\geq 2$, or GAD-2 $\geq 2$). For EMA measures, the stable baseline consisted of all available EMA responses preceding the current window, without symptom-based filtering. These reference periods captured short-term and longer-term within-person change. Both recent and stable baselines were used when available; if either could not be constructed, that comparison was omitted rather than imputed.

\textbf{Context: Patient-specific historical grounding.}
After representing the current behavioral state and its baseline-relative changes, the patient agent compares patterns across multiple time windows to establish the direction and context of the current trajectory. CHORAL accumulates these longitudinal records in a Phenotype DB, with each record containing self-reported symptom states, salient behavioral patterns, baseline-relative changes, and a patient-level summary. Each evidence item is assigned a unique identifier specifying the patient, evidence channel, week, and specific item (e.g., \texttt{[UID179-SLEEP-23\#03]}), enabling agents to cite supporting evidence directly in their reasoning. These accumulated records can later be revisited as historical context during subsequent reasoning.

\subsubsection{Domain Agents}

To address DR2, CHORAL uses separate domain agents to appraise the same shared trajectory from different domain-specific perspectives. In the formative study, clinicians noted that the same behavioral change could support multiple interpretations, indicating a need to preserve distinct appraisals before integrating them at the case level. This motivated separate domain-specific appraisal before synthesis.

We instantiate four domain agents corresponding to depression, anxiety, stress, and sleep. These domains were selected because they represent important and closely related aspects of mental health and wellbeing~\cite{clarke2014dsm,mcewen2008central} and serve as the domains through which CHORAL operationalizes multi-domain reasoning.

\textbf{Independent specialization: Domain-specific prognostic appraisal.}
Each domain agent independently evaluates the shared trajectory from the perspective of its own domain. The agent considers how the current behavioral pattern may relate to that domain, whether it raises a domain-specific prognostic concern, what evidence supports or weakens that appraisal, and what follow-up or management considerations may be relevant. For domain-specific background knowledge, the depression, anxiety, and sleep agents refer to DSM-informed criteria~\cite{american2022diagnostic}, while the stress agent draws on established stress frameworks~\cite{mcewen2008central}. This structure allows the same behavioral evidence to be interpreted separately from multiple perspectives while preserving each domain-specific appraisal before integration. The design is informed by prior medical multi-agent systems in which specialist agents with different clinical expertise independently review the same case before their appraisals are integrated~\cite{tang2024medagents,kim2024mdagents,wang2025colacare}.

\textbf{Domain Memory: Longitudinal domain-specific reasoning.}
A domain-specific prognostic appraisal is not treated as a one-time judgment. It is updated by linking an earlier appraisal with the symptom state observed afterward. Clinical reasoning similarly treats an initial interpretation as a provisional judgment that may be reviewed or revised as new evidence becomes available~\cite{nurcombe1986clinical,ilgen2016s,heneghan2009diagnostic}. Providing the full longitudinal history at every reasoning round can make prior information difficult to use consistently. Prior LLM-agent studies address this by storing past information in explicit memory and retrieving selected records during later reasoning~\cite{zhong2024memorybank,zhang2025survey}.

In CHORAL, each domain agent maintains an explicit memory of its previous domain-specific reasoning. Each memory record stores the patient's trajectory and the agent's domain-specific appraisal at that time, and is later linked with the subsequently observed symptom state and a reflection on the earlier appraisal. Selected past memories can then be retrieved in later rounds, allowing the current appraisal to be considered in relation to the patient's prior course. Domain Memory consists of two processes: \textit{Memory Construction}, which stores the result of each reasoning round and updates the previous record once the subsequent symptom state becomes available, and \textit{Memory Retrieval}, which brings selected past records into the current reasoning round.

\begin{figure}[t]
    \centering
    \includegraphics[width=0.66\linewidth]{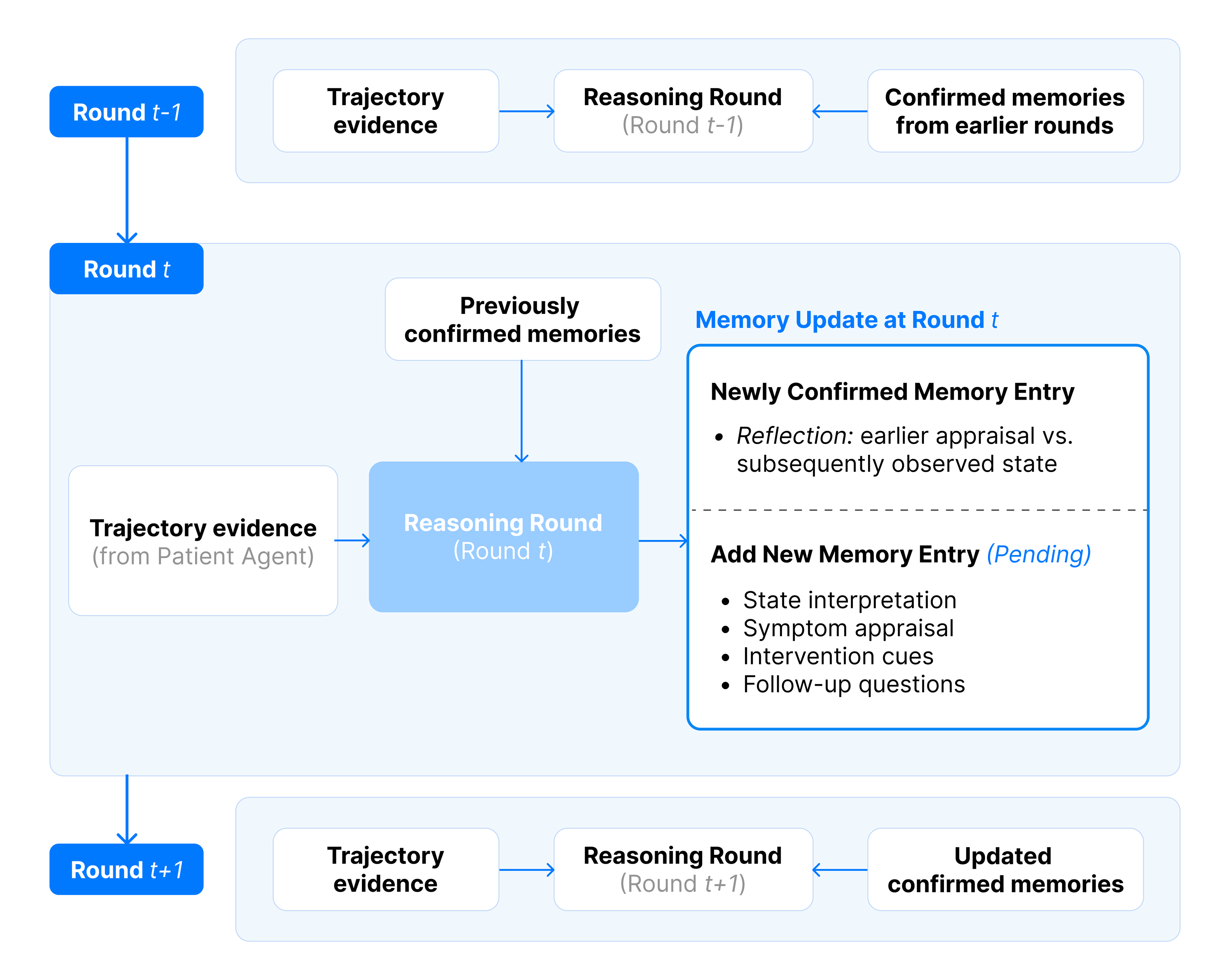}
    \caption{Domain Agent Memory across reasoning rounds. Current trajectory evidence and recent confirmed memories inform each new prognostic appraisal, which is later updated with the observed outcome and reflection.}
    \label{fig:domain-memory}
\end{figure}

\emph{Memory Construction.}
As shown in Fig.~\ref{fig:domain-memory}, each reasoning round updates previously pending memory entries whose target symptom states have become available while creating a new entry for the current round. The domain agent first interprets the shared trajectory from its own domain perspective and stores the result as a new memory entry. Each entry contains four components: a state interpretation, a domain-specific prognostic appraisal, intervention cues, and follow-up questions. The state interpretation captures how the current behavioral pattern and self-reported symptom state were understood, while the domain-specific appraisal captures the prognostic concern for that domain and its supporting evidence. Intervention cues summarize relevant management considerations, and follow-up questions identify information that should be clarified later.

Since the target symptom state is not yet available when an entry is created, the new entry is initially stored as \textit{pending}. Once the target symptom state corresponding to an earlier appraisal becomes available after the predefined prediction horizon, the earlier appraisal is compared with the observed state and a reflection is added. For example, with a two-week prediction horizon, an appraisal generated at round $t$ remains pending until the symptom state at round $t+2$ becomes available. The reflection records where the earlier judgment aligned with or differed from the subsequent course, which evidence remained useful in explaining that course, and which parts of the interpretation were less well supported. This design draws on reflection-based agent approaches that review earlier reasoning outcomes and reuse that review in later reasoning~\cite{shinn2023reflexion}.

Once the reflection is added, the corresponding entry becomes a \textit{confirmed} memory, while more recent entries whose target symptom states have not yet been observed remain pending. Over successive rounds, domain memory accumulates a linked history of patient context, domain-specific appraisal, subsequent symptom state, and reflection.

\emph{Memory Retrieval.}
At each reasoning round, the domain agent retrieves confirmed memories from earlier rounds containing its previous appraisal, the symptom state observed afterward, and the corresponding reflection. This allows the agent to consider how its earlier judgments related to the patient's subsequent course.

CHORAL retrieves recent confirmed memories rather than the entire memory history. Prior LLM-agent work similarly uses recency as an important retrieval signal~\cite{park2023generative,zhong2024memorybank}. In our implementation, each domain agent receives the three most recent confirmed memories for the current reasoning round.

\subsubsection{Information Gap Handling}

\begin{figure*}[t]
    \centering
    \includegraphics[width=0.65\textwidth]{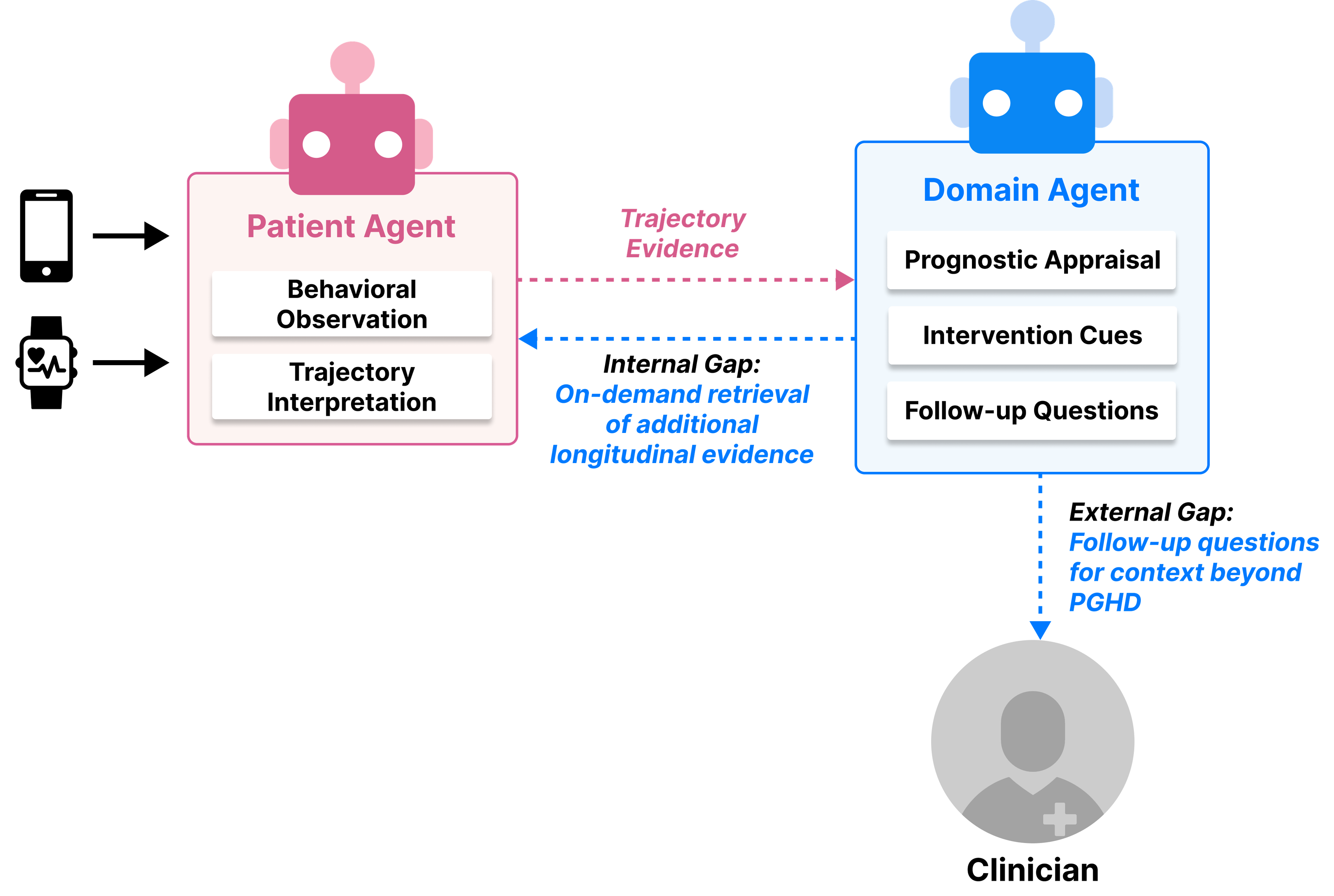}
    \caption{Overview of CHORAL's information gap handling process.}
    \label{fig:information_needs}
\end{figure*}

To operationalize DR3, CHORAL identifies what additional information is needed to clarify the initial appraisal of a domain agent and determines whether that information can be retrieved from the longitudinal patient record or should be left for clinical follow-up (Figure~\ref{fig:information_needs}). We refer to the former as an \textit{internal gap}, which can be addressed through additional retrieval from the existing longitudinal record, and the latter as an \textit{external gap}, which requires context beyond the available PGHD and is therefore retained for clinical follow-up. In the formative study, clinicians described identifying what was missing when the meaning of an observed change was unclear or did not fit well with the patient's explanation, then revisiting the patient's prior course or seeking additional context during follow-up. The clinical reasoning literature similarly treats an initial interpretation as provisional and subject to refinement as additional relevant information becomes available~\cite{kassirer1983teaching,heneghan2009diagnostic}.

\textbf{Identifying Information Needs.}
After producing an initial appraisal, each domain agent reviews which parts of its interpretation remain unclear and what additional patient-level information would help clarify them. The agent considers its interpretation together with supporting and weakening evidence, identifies the uncertainty underlying the current judgment, and specifies the information needed to reduce that uncertainty as an \textit{information need}.

These information needs are divided into two types. An \textit{internal gap} can be addressed from the existing longitudinal history, such as whether the current pattern has occurred before or how a similar behavioral change previously appeared alongside symptom states. An \textit{external gap} cannot be resolved from the current patient data, such as the situational cause of a change or the patient's subjective explanation, and is therefore left for direct assessment during subsequent clinical follow-up. This reflects selective gathering of information in clinical problem solving, where clinicians seek information relevant to the current interpretation rather than reviewing all available information again~\cite{kassirer1983teaching}.

\textbf{Query-based Retrieval and Follow-up.}
For an internal gap, the corresponding information need is converted into a \textit{system-level query} and sent to the patient agent. Rather than summarizing the entire history again, the patient agent selectively retrieves patient-level evidence related to the query from the longitudinal record and returns it to the domain agent. This follows previous retrieval-based approaches to surfacing clinically relevant information from larger patient records~\cite{alkhalaf2024applying}. For example, it may check whether the current behavioral pattern has occurred before or how a similar pattern previously co-occurred with symptom states.

For an external gap, the corresponding information need is retained as \textit{clinician-level follow-up} rather than being resolved within the system. This may include the situational context surrounding a change, the patient's subjective experience, or other information not captured in sensing or self-report data. These unresolved information needs appear as follow-up questions that clinicians can address with the patient during a subsequent visit.

The evidence returned by the patient agent is passed back to the domain agent for a refinement of the initial appraisal. The appraisal may be retained or strengthened when the evidence supports the existing interpretation, revised when the evidence points in another direction, or left uncertain when the issue remains unresolved. This reconnects the current appraisal with information already available in the longitudinal patient record and is limited to a single query--response--refinement cycle.

\subsubsection{Orchestrator}

In response to DR4, CHORAL uses an orchestrator to combine the appraisals from the depression, anxiety, stress, and sleep domain agents into an integrated assessment of the patient's overall state. In the formative study, clinicians described forming an overall judgment by comparing evidence and interpretations across domains rather than relying on a single explanation. Previous multi-agent medical systems have similarly used coordinating agents to synthesize judgments generated independently by specialist agents with different clinical perspectives~\cite{tang2024medagents,kim2024mdagents,wang2025colacare}. 

\textbf{Cross-domain Synthesis.}
The Orchestrator examines the domain-specific prognostic appraisals and supporting evidence from all four domain agents to understand how concerns and evidence across domains relate to one another. It identifies behavioral patterns or prognostic concerns supported across multiple domains while preserving interpretations that are specific to a particular domain or differ across domains. Based on these appraisals, the Orchestrator produces an \textit{initial integrated assessment} that preserves which domain provided each interpretation and what evidence supported it, thereby retaining the reasoning trace from each domain.

\textbf{Domain Review and Traceable Revision.}
After the orchestrator produces a provisional synthesis, each domain agent reviews whether its domain-specific appraisal and supporting evidence are appropriately reflected in the synthesis. The agent checks whether its interpretation has been oversimplified or integrated in a way that is inconsistent with the supporting evidence. When revision is needed, it provides feedback specifying what should be reconsidered and why. For example, it may indicate that a concern from one domain has been weakened too much or that distinct interpretations across domains have been merged in a way that removes an important difference.

The Orchestrator revises the provisional synthesis based on this feedback, after which the revised synthesis is reviewed again by the domain agents. Following prior work that uses bounded rounds of multi-agent deliberation to iteratively refine a collective assessment~\cite{kim2024mdagents}, this review-and-revision process was repeated for up to three rounds.

Once the review process is complete, the orchestrator finalizes the revised synthesis as the integrated assessment. The final result includes each domain's appraisal, major supporting evidence, areas of agreement and difference across domains, and remaining uncertainty and follow-up needs. The original domain agent appraisals and review feedback are also retained so that clinicians can trace the evidence and review process that led to the final assessment.

\textbf{Orchestrator Memory: Longitudinal integrated reasoning.}
The Orchestrator maintains a memory of how multiple domains were interpreted together in earlier reasoning rounds. After domain review and any necessary revision are completed, the final integrated assessment from each reasoning round is stored as a memory entry. Once the subsequent symptom state becomes available, the previous entry is updated to record how the earlier final integrated assessment related to the patient's later course and is then marked as \textit{confirmed}. This preserves both the earlier integrated assessment and how the corresponding cross-domain interpretation relates to the subsequent symptom course.

At a new reasoning round, the orchestrator retrieves recent confirmed memories and considers earlier final integrated assessments together with the symptom states that followed them. This allows it to consider not only the current domain agent outputs, but also cross-domain patterns that previously appeared in the same patient and how those patterns developed afterward. In our implementation, the three most recent confirmed memories are retrieved to connect the current integrated assessment with the patient's recent reasoning history.

\begin{figure*}[t]
    \centering
    \includegraphics[width=\textwidth]{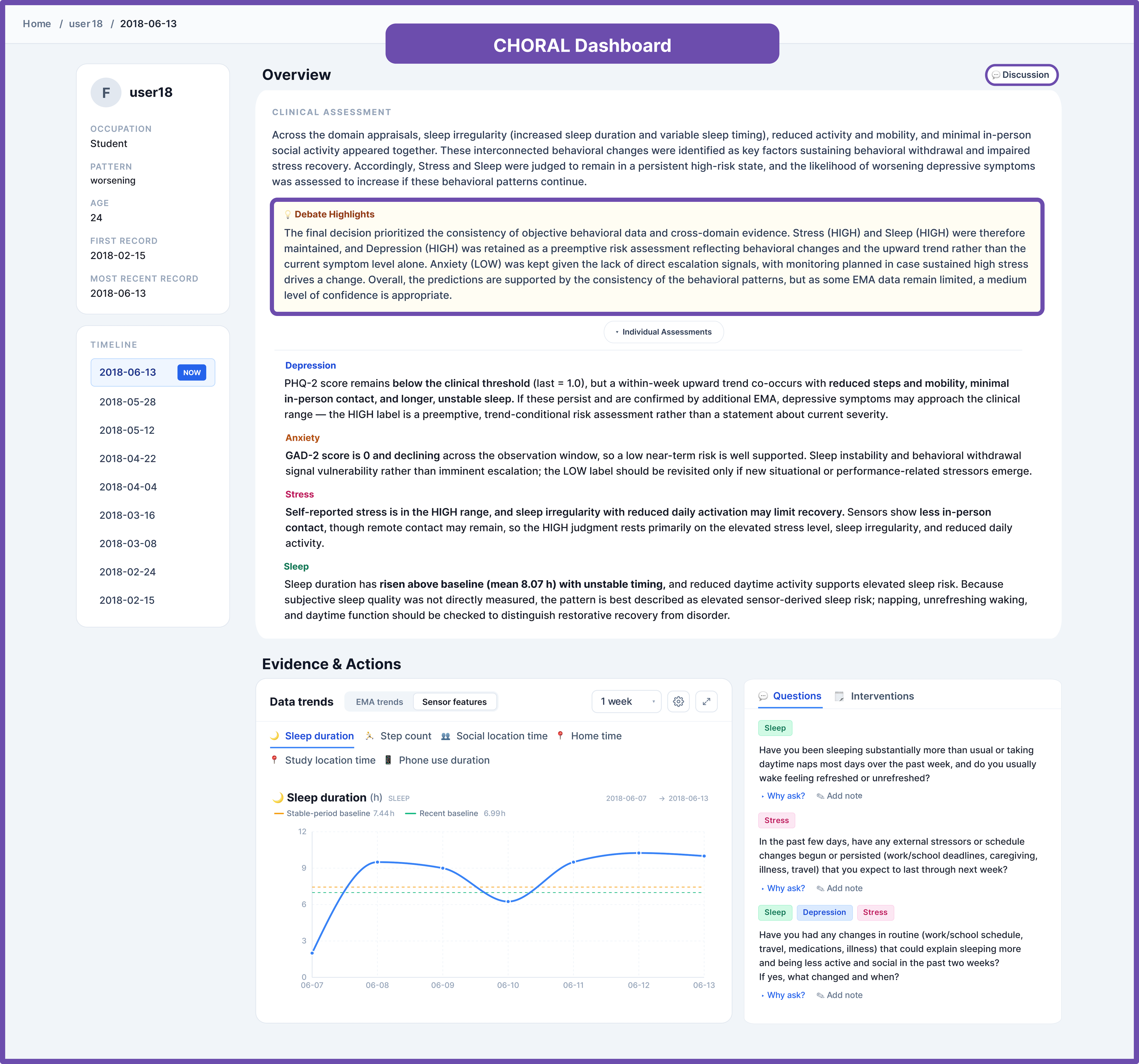}
    \caption{Overview of the CHORAL clinical dashboard. The interface prioritizes key findings, supporting evidence, and follow-up information, while providing access to Domain Agent appraisals and the Orchestrator's review process.}
    \label{fig:dashboard-overview}
\end{figure*}

\subsubsection{Clinical Dashboard}
\begin{figure*}[h]
\centering
\includegraphics[width=\textwidth]{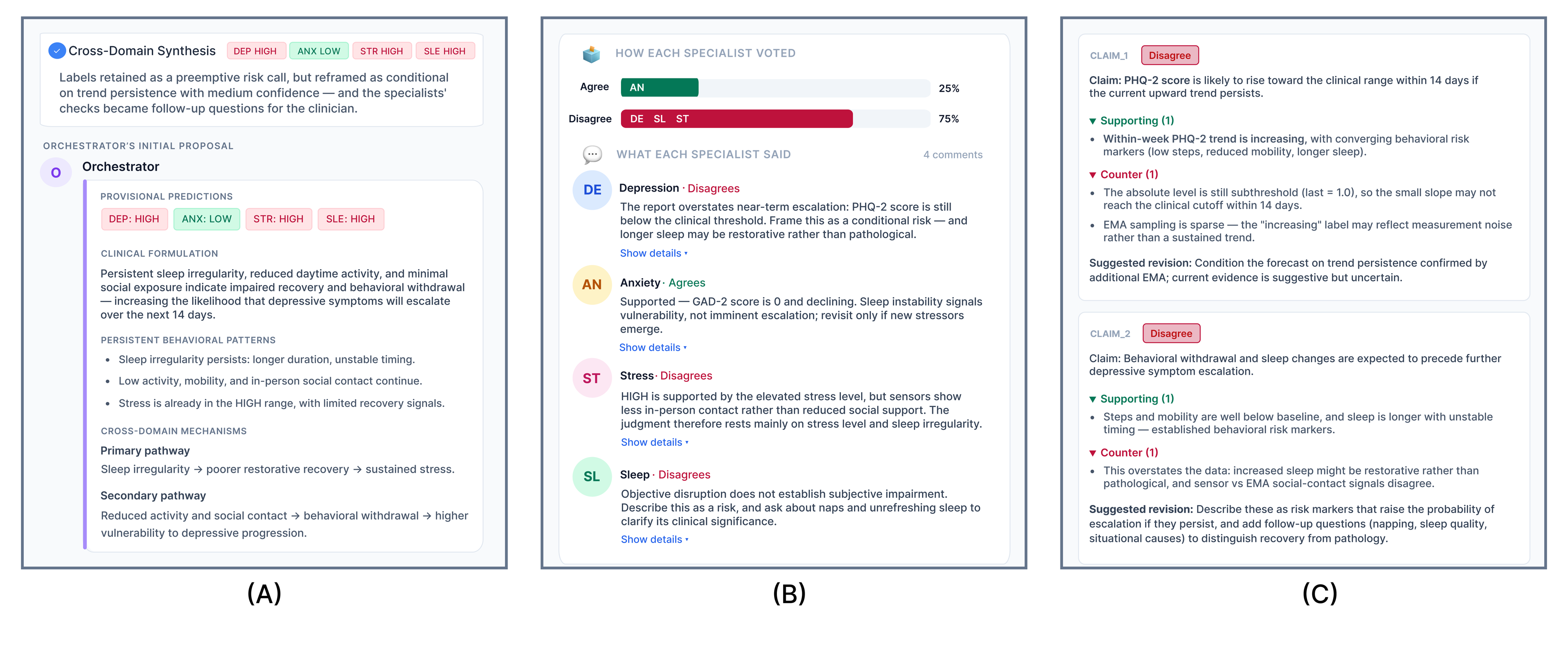}
\caption{Detail views of CHORAL's deliberation interface: (A) the Orchestrator's initial cross-domain synthesis, (B) Domain Agent review outcomes showing agreement and disagreement with the synthesis, and (C) claim-level review details showing supporting evidence, counterevidence, and suggested revisions.}
\label{fig:dashboard-discussion}
\end{figure*}
To fulfill DR5, CHORAL presents the preceding reasoning results through a clinical dashboard that helps clinicians quickly identify key findings and use them in subsequent care. In the formative study, clinicians noted that limited visit time makes it difficult to review all longitudinal data and the full reasoning trace from the beginning. CHORAL therefore prioritizes the patient's current state and recent changes, the main evidence supporting the assessment, and information relevant to follow-up, rather than presenting the entire reasoning process at once (Fig.~\ref{fig:dashboard-overview}).
For further inspection, clinicians can examine individual domain agent appraisals, the orchestrator’s review process, and the underlying reasoning trace (Fig.~\ref{fig:dashboard-discussion}). 

\section{Evaluation Methods}
We evaluated CHORAL in two ways. First, we conducted a quantitative system evaluation to examine how its design components contribute to prognostic performance and evidence grounding from longitudinal PGHD. Second, we conducted a clinician comparative study to examine how CHORAL's multi-agent orchestration supports prognostic reasoning, follow-up planning, reasoning review, and reliance support.

\subsection{Quantitative System Evaluation}
To quantitatively evaluate CHORAL, we compared the full system with reference baselines and conducted an ablation study to examine how its architectural components affect prognostic performance, evidence use, and citation validity.

\subsubsection{Dataset and Evaluation Protocol}
We evaluated CHORAL using the College Experience Study (CES) dataset~\cite{nepal2024capturing}, a longitudinal college-student cohort containing passive sensing data and repeated self-reported mental health measures. Participants completed weekly ecological momentary assessments (EMAs) of depressive symptoms using the PHQ-2, anxiety symptoms using the GAD-2, and perceived stress. These repeated self-reports were used to define the reference symptom states for quantitative evaluation.

To provide sufficient longitudinal context, we included participants with the earliest qualifying 16-week consecutive segment in which all scheduled self-reports were available and each week met the sensing-data quality criteria. Using the CES per-day quality indices~\cite{kim2026smartphone}, a day was considered complete for a sensor stream when at least 60\% of the 24 hourly intervals contained data (equivalent to at least 15 hours given the integer-valued quality indices); each week required at least five complete days for activity and at least five for location, assessed independently. After filtering, 74 participants were included in the analysis.

For each target assessment, we defined binary HIGH and LOW states for depression, anxiety, stress, and sleep. Depression and anxiety were binarized using PHQ-2 $\geq 3$ and GAD-2 $\geq 3$, respectively \cite{kroenke2003patient,kroenke2007anxiety}. For stress, we defined HIGH as scores $\geq 4$ on the 5-point scale, corresponding to responses above the neutral midpoint and indicating perceived stress.
Because CES does not include a weekly self-reported sleep measure, we derived the sleep outcome from passive sensing based on sleep duration, timing, and irregularity; accordingly, sleep results should be interpreted as prediction of a future sensor-derived sleep state rather than a self-reported symptom state. The detailed sleep-label construction is provided in Appendix~\ref{appendix-sleep-label}.

For the evaluation, the first four weeks were used to establish patient-specific baseline context. Beginning at Week 5, the system used the available longitudinal PGHD to estimate each domain's state approximately two weeks later. We selected this horizon to approximate the multi-week interval between outpatient follow-up visits; for example, measurement-based depression care commonly involves follow-up visits every 2--3 weeks during acute treatment \cite{hong2021implementing}. At each reasoning round, the system had access only to information available up to that round; future target states were not exposed until they became available for subsequent outcome confirmation. Under this two-week horizon, each appraisal remained pending until its target state became available.

The primary analysis included target assessments from Weeks 7 through 16. For example, an appraisal generated at Week 5 was evaluated against the symptom state at approximately Week 7. Across 74 participants, this yielded 740 case-weeks and 2,960 domain-level evaluation cases.

\subsubsection{Baselines and Ablations}
We compared full CHORAL against three baselines. The \emph{majority class baseline} always assigned the majority class for each domain. The \emph{persistence baseline} assumed that the patient's future target state would remain the same as the currently observed state.

We additionally constructed a \emph{single-agent baseline} as a stronger architectural comparison, preserving the shared patient agent trajectory, domain knowledge, target definitions, output requirements, and comparable longitudinal memory. After receiving the same patient agent trajectory used by CHORAL, a single LLM evaluated depression, anxiety, stress, and sleep and integrated the four domain-specific appraisals within a single generation. It therefore performed the same overall prognostic task without separate domain agents, on-demand retrieval and appraisal refinement, or orchestrator-mediated synthesis, review, and revision.

\begin{figure*}[t]
    \centering
    \includegraphics[width=\textwidth]{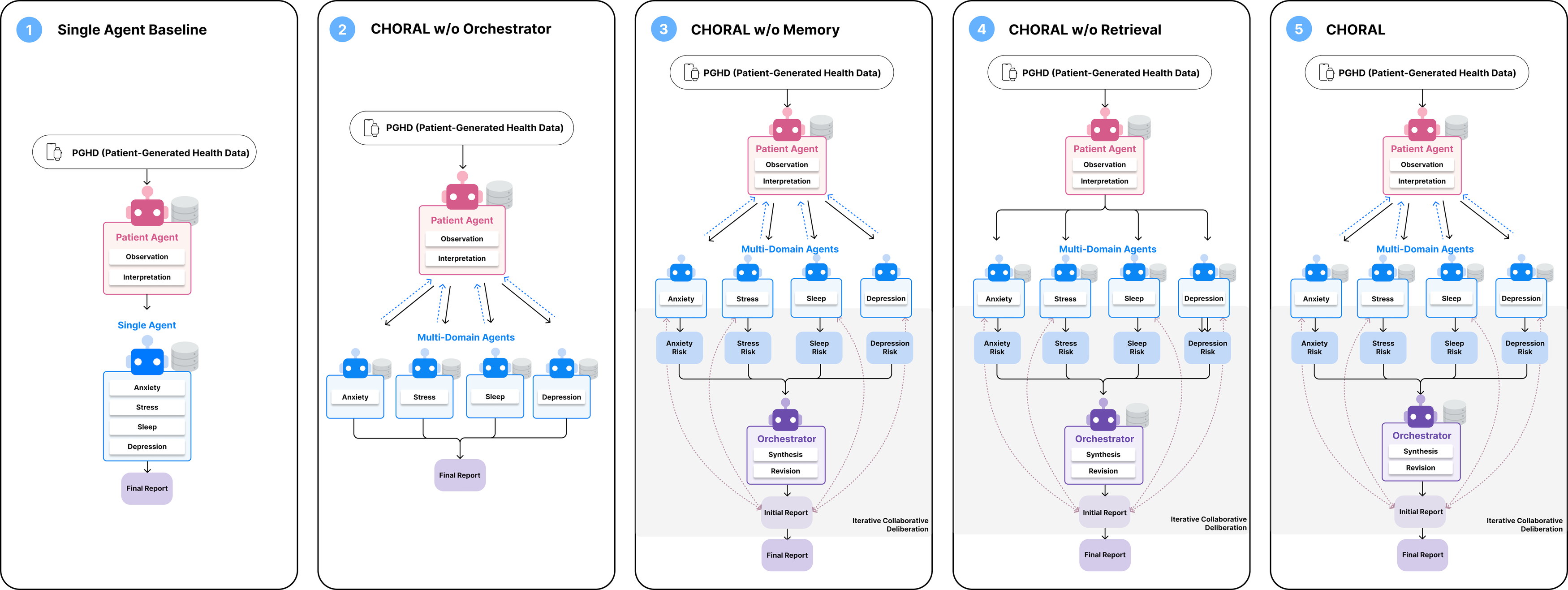}
    \caption{System configurations used for the quantitative evaluation.}
    \label{fig:ablation}
\end{figure*}

Figure~\ref{fig:ablation} summarizes the evaluated system configurations. To examine the contribution of individual CHORAL components, we evaluated three ablated variants:

\begin{itemize}
    \item \textbf{w/o On-demand Retrieval}: removes the mechanism through which a domain agent requests additional evidence from the longitudinal patient record
    \item \textbf{w/o Memory}: removes domain memory and orchestrator memory, while retaining the shared patient agent trajectory and longitudinal patient evidence used across all conditions
    \item \textbf{w/o Orchestrator}: retains the domain-specific prognostic appraisals but removes cross-domain synthesis, domain review, and conditional revision by the Orchestrator
\end{itemize}

\subsubsection{Evaluation Metrics}
We evaluated CHORAL along two dimensions: prognostic performance and evidence grounding. For prognostic performance, balanced accuracy and deterioration detection were complementary primary metrics. Balanced accuracy accounts for class imbalance across HIGH and LOW states, whereas deterioration detection captures the ability to anticipate LOW-to-HIGH transitions~\cite{dormechele2026digital}. Precision, recall, and F1 were reported as complementary metrics, with HIGH as the positive class. Balanced accuracy, precision, recall, and F1 were macro-averaged across the four domains, whereas deterioration detection pooled all observed LOW-to-HIGH transitions. For currently LOW cases, an \emph{alert} was a predicted future HIGH state; \emph{alert precision} was the proportion of alerts followed by a LOW-to-HIGH transition, and \emph{false-alert rate} was the proportion of LOW-to-LOW cases incorrectly predicted HIGH. For key comparisons, 95\% confidence intervals were estimated from 2,000 bootstrap resamples at the participant level, preserving all case-weeks and domain predictions within each sampled participant.

For evidence grounding, we evaluated the final output for each condition: the integrated assessment for orchestrated conditions, concatenated final domain-agent appraisals for w/o Orchestrator, and the final assessment for the single-agent baseline. \emph{Evidence channel coverage} was the proportion of available (non-empty) channels, drawn from seven channel types (ACT, EMA, LOC, PHONE, SIG, SLEEP, and SOCIAL), that were cited at least once per case-week; each channel was counted at most once. \emph{Invalid evidence citation rate} was the proportion of citation occurrences that did not exactly match an evidence identifier in the corresponding patient record, with repeated tags counted separately. Counts were pooled across cases, and invalid citations were manually inspected. These metrics capture reference breadth and identifier validity, not clinical relevance or semantic correctness.

\subsection{Clinician Evaluation}
We conducted a user study with clinicians to evaluate how CHORAL supports prognostic reasoning, follow-up planning, reasoning review, and reliance support in comparison with a single-agent approach.

\subsubsection{Study Design}\label{main-study-design}
We used a within-subject design comparing a single-agent condition with CHORAL. The single agent received the same Patient Agent trajectory and produced four domain appraisals and an integrated assessment in a single generation. In addition to separate domain agents, CHORAL included on-demand evidence retrieval and appraisal refinement, orchestrator-mediated synthesis, review, and revision, and intermediate deliberation views that were not available in the single-agent condition. Thus, the clinician study compared the two system configurations as a whole rather than isolating the effect of agent organization.

We selected four cases from the CES evaluation set to represent distinct longitudinal courses: \emph{deterioration}, \emph{recovery}, \emph{chronic high}, and \emph{mixed trajectory}. Cases were selected to capture variation in longitudinal trajectory patterns rather than to favor a particular system outcome. Participants reviewed only information available through the corresponding case-week; subsequent target outcomes were not shown during the evaluation.

\subsubsection{Participants}
We recruited 22 psychiatry clinicians, including 14 psychiatry residents and eight board-certified psychiatrists (7 female; age $M$=34.5, $SD$=6.4). Clinical experience was defined as total training and practice experience, including internship, residency, and subsequent clinical practice. Eight participants had 1--3 years of experience, six had 4--5 years, three had 6--10 years, and five had more than 10 years of experience. Full participant demographics are provided in Appendix~\ref{appendix-participants}. Participants received approximately USD 150 as compensation. The study protocol was approved by the IRB, and all participants provided informed consent before participation. 

\subsubsection{Procedure} 
After a brief familiarization with the interface, each participant reviewed four pre-generated patient case reports, two in the single-agent condition and two in the CHORAL condition. Reports were formatted for presentation consistency, but their substantive reasoning and prognostic content were not manually edited. Case-to-condition assignments and presentation order were balanced across participants, yielding 11 evaluations per case in each condition. After each case, participants completed the corresponding questionnaire; after all four cases, they took part in a semi-structured interview comparing the two conditions. Each session lasted approximately 40--50 minutes and was conducted online.

\subsubsection{Measures}
The questionnaire assessed two groups of outcomes corresponding to RQ2 and RQ3. For \textit{Clinical Usefulness, Clinical Actionability, and Workflow Fit} (RQ2), questionnaire items assessed the clinical usefulness of the prognostic judgment, the usefulness of integrating multiple domains, support for identifying information requiring further confirmation and prioritizing follow-up concerns, and the suitability of the presented information for clinical workflow. For \textit{Contestability and Reliance Support} (RQ3), questionnaire items assessed clinicians' perceived support for tracing the final assessment to its supporting evidence and domain-specific prognostic appraisals, identifying disagreements or alternative interpretations across domains, determining which parts of the reasoning required further review, and distinguishing between parts of the system output they could rely on and those requiring further verification. The full questionnaire is provided in Appendix~\ref{appendix-questionnaire}.

Semi-structured interviews explored how the two conditions supported clinicians' understanding and review of prognostic judgments. We examined how information and reasoning traces supported clinicians in questioning the assessment, identifying information gaps and follow-up priorities, and determining which outputs were useful, burdensome, or trustworthy and which required further verification.

\subsubsection{Analysis}
Because questionnaire responses were collected on a 7-point Likert scale, we used Wilcoxon signed-rank tests to compare the two conditions. For each participant, ratings were averaged across the two cases assigned to each condition, yielding one participant-level score per condition for each item. We report matched-pairs rank-biserial correlations ($r_{rb}$) as effect sizes. For multiple comparisons, item-level $p$-values were corrected using the Benjamini--Hochberg procedure at an FDR of $q=.05$, separately within the RQ2 and RQ3 measure families.

Interview transcripts were qualitatively coded by two researchers, who iteratively compared and refined the codes and recurring themes through discussion.
\section{Results}

\subsection{RQ1. Architectural Contributions}
\begin{table}[t]
\centering
\caption{Prognostic performance of CHORAL compared with reference baselines.}
\label{tab:baseline-performance}
\begin{tabular}{lccccc}
\toprule
\textbf{Condition} & \textbf{BalAcc} & \textbf{Precision} & \textbf{Recall} & \textbf{F1} & \textbf{Deterioration Detection} \\
\midrule
Majority Baseline
& 0.500
& 0.000
& 0.000
& 0.000
& 0/295 (0\%) \\

Persistence Baseline
& 0.672
& \textbf{0.486}
& 0.478
& 0.482
& 0/295 (0\%) \\

Single Agent Baseline
& 0.639
& 0.351
& 0.558
& 0.430
& 104/295 (35\%) \\

\textbf{CHORAL}
& \textbf{0.673}
& 0.403
& \textbf{0.612}
& \textbf{0.484}
& \textbf{118/295 (40\%)} \\
\bottomrule
\end{tabular}
\end{table}

\subsubsection{Prognostic Performance} 
CHORAL achieved a balanced accuracy of 0.673, comparable to persistence (0.672) and higher than the single-agent (0.639) and majority (0.500) baselines (Table~2). The difference was 0.034 versus the single-agent baseline (95\% CI [0.008, 0.058]) and 0.001 versus persistence (95\% CI [-0.021, 0.026]). Persistence detected none of the 295 LOW-to-HIGH transitions by construction, whereas CHORAL detected 118 (40.0\%) and the single-agent baseline 104 (35.3\%). Relative to persistence, CHORAL had higher recall (0.612 vs. 0.478) but lower precision (0.403 vs. 0.486).

\begin{table}[t]
\centering
\caption{Ablation results for CHORAL's architectural components. Deterioration indicates detection of LOW-to-HIGH transitions.}
\label{tab:ablation-performance}
\begin{tabular}{lccccc}
\toprule
\textbf{Condition} & \textbf{BalAcc} & \textbf{Precision} & \textbf{Recall} & \textbf{F1} & \textbf{Deterioration Detection} \\
\midrule

w/o Orchestrator
& 0.657
& 0.353
& \textbf{0.638}
& 0.454
& \textbf{137/295 (46\%)} \\

w/o Memory
& 0.641
& 0.393
& 0.571
& 0.461
& 103/295 (35\%) \\

w/o On-demand Retrieval
& 0.640
& 0.354
& 0.560
& 0.433
& 129/295 (44\%) \\

\textbf{CHORAL}
& \textbf{0.673}
& \textbf{0.403}
& 0.612
& \textbf{0.484}
& 118/295 (40\%) \\

\bottomrule
\end{tabular}
\end{table}

Ablations showed complementary trade-offs (Table~\ref{tab:ablation-performance}). Removing on-demand retrieval or memory reduced balanced accuracy to 0.640 and 0.641, respectively. Removing the Orchestrator lowered precision (0.403 to 0.353) but increased recall (0.612 to 0.638). 
Although w/o Orchestrator and w/o On-demand Retrieval showed higher deterioration sensitivity (46\% and 44\% vs. 40\%), CHORAL achieved the highest alert precision (29\%) and the lowest false-alert rate (14.4\%; Table~\ref{tab:alert-performance}). Thus, CHORAL's contribution is not uniformly higher sensitivity, but a different operating trade-off: the full system detected fewer deterioration cases than some ablations while producing fewer false alerts and higher overall balanced accuracy.

\begin{table}[t]
\centering
\caption{Alert-level performance for currently LOW cases ($n=2{,}342$). 
Alerts denote predicted future HIGH states; persistence is omitted because it raises none by construction. 
False-alert rate is computed over LOW-to-LOW cases ($n=2{,}047$).}
\label{tab:alert-performance}
\small
\setlength{\tabcolsep}{4pt}
\begin{tabular}{lccc}
\toprule
\textbf{Condition}
& \shortstack{\textbf{Deterioration}\\\textbf{Sensitivity}}
& \shortstack{\textbf{Alert}\\\textbf{Precision}}
& \shortstack{\textbf{False-Alert}\\\textbf{Rate}} \\
\midrule
Single-Agent Baseline           & 35\% & 22\% & 17.7\% \\
w/o Orchestrator                & 46\% & 24\% & 20.7\% \\
w/o Memory                      & 35\% & 25\% & 15.4\% \\
w/o On-demand Retrieval         & 44\% & 25\% & 19.1\% \\
\textbf{CHORAL}                 & 40\% & \textbf{29\%} & \textbf{14.4\%} \\
\bottomrule
\end{tabular}
\end{table}

\begin{figure*}[t]
    \centering
    \includegraphics[width=0.9\textwidth]{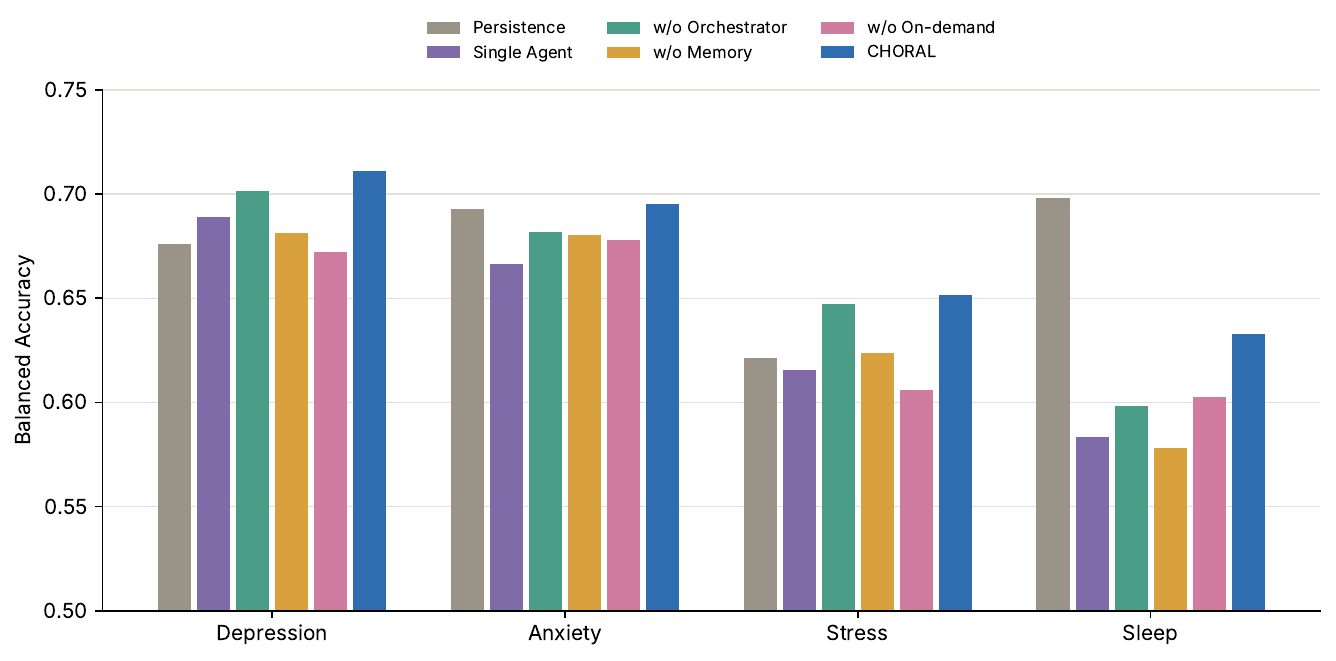}
    \caption{Balanced accuracy across mental-health domains for the persistence baseline, single-agent baseline, CHORAL ablations, and full CHORAL.}
    \label{fig:domain-balacc}
\end{figure*}

Across domains (Figure~\ref{fig:domain-balacc}), CHORAL achieved the highest balanced accuracy for depression (0.711), anxiety (0.695), and stress (0.651). Persistence outperformed CHORAL only for sleep (0.698 vs. 0.633), where HIGH states were more prevalent (42\% vs. 10--21\%) and persistent, with only 34\% transitioning to LOW versus 54--60\% in the other domains. CHORAL's higher recall but lower specificity therefore reduced sleep balanced accuracy. Deterioration detection also varied by domain (21--58\%), indicating heterogeneity behind the pooled 40\% result. Detailed domain-level results are reported in Appendix~\ref{appendix-domain-performance}.

\subsubsection{Evidence Use and Citation Validity} 
\begin{table}[t]
\centering
\small
\caption{Evidence grounding across the single-agent baseline, CHORAL ablations, and full CHORAL.}
\label{tab:evidence-grounding}
\begin{tabular}{lcc}
\toprule
\textbf{Condition} &
\textbf{Evidence Channel Coverage} &
\textbf{Invalid Evidence Citation Rate} \\
\midrule

Single Agent Baseline
& 70.1\%
& 0.0112\% \\

 w/o Orchestrator
& 93.8\%
& 0.0035\% \\

 w/o Memory
& 88.7\%
& 0.0000\% \\

 w/o On-demand Retrieval
& 87.4\%
& 0.0000\% \\

CHORAL
& 89.2\%
& 0.0031\% \\

\bottomrule
\end{tabular}
\end{table}

The multi-agent conditions referenced a broader range of patient evidence than the single-agent baseline (Table~\ref{tab:evidence-grounding}). Evidence channel coverage was 70.1\% for the single-agent baseline and 87.4--93.8\% across the multi-agent conditions, with full CHORAL reaching 89.2\%. Coverage was highest in the w/o Orchestrator condition (93.8\%), suggesting that domain-level appraisal referenced the broadest range of evidence, whereas orchestration produced a more selective synthesis.

Invalid evidence citation rates were at most 0.0112\% across conditions. Manual inspection revealed two distinct types of errors. In CHORAL and the w/o Orchestrator condition, the agent abbreviated the \texttt{SOCIAL} channel as \texttt{SOC}; however, the cited social-exposure evidence itself existed in the patient record and was reported correctly. In the single-agent baseline, the cited channel and week were valid, but the specific item number did not exist in that record (e.g., \texttt{[UID179-SLEEP-23\#10]} when the SLEEP channel contained only items \texttt{\#01}--\texttt{\#04}). These citations therefore referred to non-existent evidence items, whereas the CHORAL errors still pointed to identifiable evidence in the patient record.

\subsection{RQ2. Clinical Usefulness, Clinical Actionability, and Workflow Fit}
\begin{table*}[t]
\centering
\small
\caption{Clinician ratings of clinical usefulness, actionability, and workflow fit for the single-agent and CHORAL.}
\label{tab:rq2_clinician_eval}

\begin{tabular}{p{0.34\textwidth}ccccccc}
\toprule
\textbf{Questionnaire Item} &
\textbf{Single Agent M (SD)} &
\textbf{CHORAL M (SD)} &
\textbf{$\Delta$} &
\textbf{$W$} &
\textbf{$p$} &
\textbf{$q$} &
\textbf{$r_{rb}$} \\
\midrule

\multicolumn{8}{l}{\textit{\textbf{Clinical Usefulness}}} \\

1. Clinically useful given recent changes
& 4.75 (0.84)
& 5.07 (0.94)
& +0.32
& 49.0
& .103
& .190
& +0.43 \\

2. Evidence supported clinical appraisal
& 4.75 (0.86)
& \textbf{5.20 (0.84)}
& \textbf{+0.45}
& \textbf{11.0}
& \textbf{.004}
& \textbf{.028}
& \textbf{+0.82} \\

3. Cross-domain aided interpretation
& 4.86 (1.00)
& 5.30 (1.09)
& +0.43
& 40.5
& .024
& .083
& +0.57 \\

\midrule
\multicolumn{8}{l}{\textit{\textbf{Clinical Actionability}}} \\

4. Identified what needs further verification
& 4.57 (0.98)
& 4.84 (1.17)
& +0.27
& 27.5
& .108
& .190
& +0.48 \\

5. Aided prioritization of prognostic concerns
& 4.95 (0.95)
& 5.02 (0.99)
& +0.07
& 57.0
& .561
& .561
& +0.16 \\

\midrule
\multicolumn{8}{l}{\textit{\textbf{Workflow Fit}}} \\

6. Amount and organization of information
& 4.70 (1.10)
& 4.50 (1.18)
& -0.20
& 52.5
& .248
& .347
& -0.31 \\

7. Integrable into existing workflow
& 4.57 (1.13)
& 4.70 (1.20)
& +0.14
& 29.5
& .449
& .524
& +0.24 \\

\bottomrule
\end{tabular}
\end{table*}

\subsubsection{Clinical Usefulness}
For clinical usefulness, CHORAL was generally rated more highly than the single-agent. Evidence-supported clinical appraisal showed the largest and only statistically significant difference after Benjamini--Hochberg correction ($\Delta=+0.45$, $p=.004$, $q=.028$, $r_{rb}=.82$), while cross-domain integration also received a higher rating but did not remain significant after correction ($\Delta=+0.43$, $p=.024$, $q=.083$).

Participants described CHORAL as making the basis of the final prognostic judgment easier to review. Although the single-agent presented domain-level findings and a final assessment, clinicians sometimes had to infer how those findings contributed to the final judgment. In CHORAL, the synthesis and review process made this connection more explicit. P6 described this as a ``gap in between'' in the single-agent condition: ``With the single-agent version, I could see the individual domain assessments and the final conclusion, but there was still a gap in between. With CHORAL, I could follow which concerns were carried forward, which were toned down, and where the interpretation changed.''

Participants also valued having access to supporting findings that were not emphasized in the final summary. These details gave them additional context for reviewing the final assessment rather than relying only on the system's main conclusion. P15 noted that ``the final summary focuses on the main concern, but CHORAL still showed the findings that were not emphasized,'' which helped them understand ``what had been considered but left in the background.'' P9 similarly emphasized that they were ``not going to take the AI's final judgment at face value anyway,'' and that the ``smaller clues and contextual details were sometimes more useful than the conclusion itself'' when forming their own view of the patient.

\subsubsection{Clinical Actionability}
For clinical actionability, CHORAL received slightly higher ratings than the single-agent, but neither difference was statistically significant.

Participants described CHORAL as useful when the available data left a case difficult to judge, providing an additional perspective similar to a second opinion. P17 described using it ``almost like a second opinion,'' while P12 noted that it prompted reconsideration of whether they had ``overlooked another way of interpreting the case.''

At the same time, participants noted that identifying issues requiring further attention did not necessarily determine follow-up priorities. Prioritization depended on additional clinical context, including the patient's chief complaint, treatment and medication history, and recent life events. P19 observed that ``the system can tell me what may need more attention, but it cannot really decide what should come first.'' P14 similarly emphasized that even when two patients showed similar patterns in the data, what they would focus on first could differ depending on why the patient came in and what had been happening in treatment.

Participants also found that unresolved disagreements could help indicate what still needed to be checked during follow-up. P11 explained, ``The disagreement pointed me to the part of the prognosis that was still uncertain. Rather than just accepting the final assessment, it gave me a reason to look for additional context or ask the patient something more specific.''

However, translating these identified concerns into concrete follow-up questions remained challenging. Although the suggested questions helped indicate what might need to be explored, they were not always sufficiently tailored for direct use with patients. Participants noted that the wording and delivery would need to be adapted to the individual patient and clinical context, such as the patient's level of understanding, age-related communication preferences, and primary concerns. P3 described the questions as useful for reminding them what to cover but ``fairly generic,'' noting that they would phrase them differently depending on the patient. P20 similarly added that some questions would need to be simplified or made more direct depending on the patient.

\subsubsection{Workflow Fit}
For workflow fit, differences between the two conditions were small. CHORAL received a slightly lower rating for the amount and organization of information ($\Delta=-0.20$, $q=.347$), while ratings for integration into existing workflow were similar between conditions ($\Delta=+0.14$, $q=.524$).

CHORAL was viewed as more useful before a visit than during the encounter itself. P8 noted that it could help them ``quickly see what has changed and what I should pay attention to before going into the visit.''

During the visit, however, participants noted that the amount of information could become burdensome because clinicians already need to attend to the EMR, medications, prior notes, and the ongoing conversation with the patient. P14 said they ``probably wouldn't read the whole deliberation from top to bottom,'' and P2 similarly noted that several additional paragraphs of reasoning could make it difficult to identify the main point.

Participants therefore preferred selective access to the detailed reasoning rather than reviewing the full coordination process each time. They were particularly interested in points where agents disagreed, where the assessment changed after review, or where uncertainty remained unresolved. P15 summarized this preference as wanting to see ``where they disagreed, what changed, and what is still unresolved.''

\subsection{RQ3. Contestability and Reliance Support}
\begin{table*}[t]
\centering
\small
\caption{Clinician ratings of contestability and reliance support for the single-agent and CHORAL.}
\label{tab:rq3_clinician_eval}

\begin{tabular}{p{0.34\textwidth}ccccccc}
\toprule
\textbf{Questionnaire Item} &
\textbf{Single Agent M (SD)} &
\textbf{CHORAL M (SD)} &
\textbf{$\Delta$} &
\textbf{$W$} &
\textbf{$p$} &
\textbf{$q$} &
\textbf{$r_{rb}$} \\
\midrule

\multicolumn{8}{l}{\textit{\textbf{Contestability Support}}} \\

1. Traceability of supporting evidence
& 5.20 (0.57)
& 5.55 (0.79)
& +0.34
& 45.0
& .071
& .086
& +0.47 \\

2. Identifying alternative interpretations across mental-health domains
& 4.25 (1.15)
& \textbf{5.41 (0.91)}
& \textbf{+1.16}
& \textbf{7.5}
& \textbf{.001}
& \textbf{.006}
& \textbf{+0.90} \\

3. Locating grounds on which to challenge
& 4.45 (1.18)
& \textbf{5.07 (1.13)}
& \textbf{+0.61}
& \textbf{21.5}
& \textbf{.003}
& \textbf{.008}
& \textbf{+0.77} \\

\midrule
\multicolumn{8}{l}{\textit{\textbf{Reliance and Verification Support}}} \\

4. Reliance as a clinical reference
& 4.34 (0.97)
& 4.68 (1.01)
& +0.34
& 33.0
& .064
& .086
& +0.51 \\

5. Distinguishing trust from verification
& 4.61 (0.94)
& \textbf{5.11 (0.79)}
& \textbf{+0.50}
& \textbf{24.0}
& \textbf{.021}
& \textbf{.042}
& \textbf{+0.65} \\

6. Low risk of inappropriate judgment
& 5.32 (0.80)
& 5.43 (0.99)
& +0.11
& 17.0
& .265
& .265
& +0.38 \\

\bottomrule
\end{tabular}
\end{table*}
RQ3 examined clinicians' perceptions of contestability and reliance support (Table~\ref{tab:rq3_clinician_eval}).

\subsubsection{Contestability Support}
Compared with the single-agent, CHORAL received significantly higher ratings for helping clinicians identify alternative interpretations across mental-health domains ($\Delta=+1.16$, $p=.001$, $q=.006$, $r_{rb}=.90$) and for helping them locate grounds on which to challenge the system's judgment ($\Delta=+0.61$, $p=.003$, $q=.008$, $r_{rb}=.77$).

Participants noted that, although the single-agent presented domain-level findings and a final judgment, they still had to work out for themselves which parts of the assessment they disagreed with or wanted to examine more closely. P14 explained, ``With the single-agent version, I could see the final judgment, but if I disagreed with it, I still had to work out for myself which part of the reasoning I actually disagreed with.'' In CHORAL, disagreements and alternative interpretations raised by domain agents remained visible, making it easier for clinicians to see where the final assessment could be interpreted differently. P4 noted, ``When a domain agent disagreed with the orchestrator, it made me question whether the final assessment was capturing the whole picture. I would want to know what was driving that difference before feeling comfortable with the prognosis.''

Participants also emphasized that the value of disagreement came from the reasoning and evidence behind it, rather than from disagreement alone. P1 explained, ``The agreement itself was not that important to me. What mattered was the reasoning behind it. If an agent said, `I agree because of this pattern,' or `I disagree because this finding points another way,' that gave me something more to work with clinically.''

Clinicians then used these alternative interpretations to examine the underlying patient data and assess which interpretation was better supported. P16 said, ``When I saw that the domain agent had a different interpretation, I went back to the underlying patient data to see what each judgment was based on. That helped me decide which interpretation was better supported.'' This also gave clinicians more concrete grounds for challenging the final judgment. P18 described this as giving them ``something concrete to challenge,'' explaining, ``Instead of just thinking, `I'm not sure I agree with this,' I could say, `I think this interpretation is too strong for the evidence we have.'{}''

\subsubsection{Reliance and Verification Support}
CHORAL received a significantly higher rating for supporting clinicians in distinguishing between aspects of the system output they could rely on and those requiring further verification ($\Delta=+0.50$, $p=.021$, $q=.042$, $r_{rb}=.65$). Reliance on the system as a clinical reference was also rated higher under CHORAL, but the difference did not reach statistical significance ($\Delta=+0.34$, $p=.064$, $q=.086$).

Participants described CHORAL as helping them decide which parts of the system's reasoning they could rely on and which required further verification. Rather than simply accepting the final conclusion, they compared the intermediate reasoning with their own clinical judgment. P7 explained, ``Seeing the intermediate reasoning did not make me trust everything it said. I could compare it with my own thinking and decide which parts I was comfortable relying on and which parts I still wanted to verify.'' P17 similarly noted, ``I would not say I trusted the whole assessment equally.''

Participants also noted that seeing multiple interpretations and the subsequent coordination process contributed to how credible the final assessment felt. P5 said, ``I trusted it more because it did not look like the system had jumped straight to one answer. I could see that different explanations had actually been considered before it reached the final assessment.'' P12 similarly emphasized that the system ``did not hide the uncertainty'' and showed that ``there were several possible ways to read the same pattern before settling on the final interpretation.''

Even when they did not inspect every reasoning step, some participants felt more comfortable using CHORAL as a clinical reference because they could see that the final assessment had been reviewed from multiple perspectives. P10 summarized this view: ``I probably would not read every exchange between the agents, but just knowing that the final answer was checked from several angles makes me a little more comfortable using it.''

\section{Discussion}

\subsection{Multi-Agent Reasoning for Longitudinal PGHD}
CHORAL achieved balanced accuracy comparable to persistence while detecting 40\% of observed LOW-to-HIGH transitions. Because persistence carries forward the current state by construction, its inability to anticipate such transitions is structural rather than evidence of uniformly inferior performance. CHORAL also showed higher recall but lower precision than persistence, highlighting a trade-off between identifying future HIGH states and avoiding false positives. Compared with the single-agent baseline, CHORAL achieved higher balanced accuracy and broader evidence use. This broader evidence use is particularly relevant in mental health sensing, where behavioral signals often require contextual interpretation rather than direct mapping to symptom labels~\cite{englhardt2024classification}. Related work has similarly argued that clinical sensing systems should be evaluated not only by predictive performance, but also by the additional information and actionable insight they provide for care~\cite{adler2024beyond}. From this perspective, CHORAL's results suggest value in incorporating broader patient evidence and seeking additional context when needed, rather than relying on predictive performance alone.

The ablation results further suggest that CHORAL's additional reasoning stages were not simply redundant. Removing on-demand retrieval produced the largest decreases in balanced accuracy, recall, and F1, consistent with additional evidence helping when the initially available information was insufficient. Removing memory also reduced balanced accuracy and recall, suggesting a role for patient-specific historical context. In contrast, removing the orchestrator increased recall but reduced precision, while also increasing Evidence Channel Coverage. This pattern is consistent with orchestration serving less to expand the amount of evidence considered and more to compare domain-level judgments and moderate overly strong HIGH predictions. Taken together, retrieval, memory, and orchestration supported complementary functions: additional evidence, historical context, and judgment adjustment. Thus, CHORAL's value lies less in using multiple agents per se and more in structuring longitudinal PGHD reasoning around complementary information functions. This is consistent with prior medical multi-agent work on specialist roles and coordination \cite{kim2024mdagents,chen2025mdteamgpt,wang2025colacare} and with health-data sensemaking research showing that longitudinal interpretation often requires combining multiple data sources, relating current observations to prior context, and seeking additional information when observed patterns are not fully explained \cite{raj2019clinical}.

However, richer reasoning did not yield the highest performance in every mental-health domain. CHORAL outperformed the persistence baseline for depression, anxiety, and stress, whereas persistence achieved higher balanced accuracy for sleep. In our data, HIGH sleep states were also more likely than HIGH states in the other mental-health domains to remain HIGH at the subsequent assessment, suggesting that the recent sleep state itself already provided a relatively strong signal about the future state. Thus, the relative benefit of additional reasoning may depend in part on how strongly the target state persists over time, and the same level of multi-agent reasoning may not be necessary across all domains. This observation is consistent with adaptive multi-agent approaches that vary collaboration structure according to task complexity \cite{kim2024mdagents}. Future systems could therefore explore allocating deeper reasoning selectively when additional evidence, historical context, or cross-domain coordination may be needed.

\subsection{Multi-Agent Deliberation for Clinician Review}
In the clinician evaluation, CHORAL received significantly higher ratings than the single-agent condition for helping clinicians identify alternative interpretations and find grounds to challenge the system's judgment. The largest difference between the two conditions was in clinicians' ratings of support for recognizing alternative interpretations. Qualitative findings further showed that clinicians valued disagreements between domain-agent assessments and the orchestrator's integrated judgment, particularly when the reasoning and evidence behind those disagreements were visible. These disagreements prompted clinicians to revisit the corresponding patient evidence, compare competing interpretations, and identify specific parts of the final judgment that warranted closer examination. Thus, CHORAL's deliberation provided concrete review points through which clinicians could examine and challenge the system's prognostic judgment. Because the clinician study compared complete system configurations, however, these differences cannot be attributed to agent organization alone and may reflect the combined effects of CHORAL's reasoning, retrieval, review, and presentation mechanisms.

This finding extends prior clinical XAI research showing that clinicians use explanations to compare AI outputs against their own judgments and verify their basis \cite{tonekaboni2019clinicians,sivaraman2023ignore}. Clinicians may selectively accept, modify, or defer aspects of an AI recommendation rather than treating it as a whole \cite{sivaraman2023ignore}. Work on contestable AI likewise emphasizes access to information that enables users to interrogate and challenge AI decisions \cite{ploug2020four}. CHORAL's alternative interpretations and cross-agent disagreements support this process by exposing where and why a judgment may warrant challenge rather than presenting only a consolidated conclusion.

CHORAL received a significantly higher rating for supporting clinicians in distinguishing what to rely on from what required further verification. Although reliance on the system as a clinical reference was also rated higher under CHORAL, this difference did not reach statistical significance, and ratings for the risk of inappropriate judgment did not differ between conditions. These findings indicate greater perceived support for selective reliance and verification, rather than demonstrating that clinicians' actual reliance was calibrated to the correctness of system outputs. This interpretation is consistent with work in human--AI decision making that frames the goal not as maximizing trust, but as supporting appropriate reliance, whereby users rely on AI when appropriate and verify or override it when necessary \cite{cao2024designing,ma2025towards}. Research on health AI likewise suggests that trust can be selectively formed depending on the recommendation and decision context \cite{van2025selective}. At the same time, P10's account suggests that visible deliberation can provide reassurance even without detailed inspection. Because CHORAL's agents share the same backbone model, agreement among agents should not be interpreted as independent clinical corroboration.

These findings suggest that clinician-facing multi-agent systems should preserve deliberation that meaningfully supports review, particularly competing interpretations, consequential disagreements, and unresolved uncertainty. Rather than exposing every agent interaction, systems may benefit from prioritizing reasoning events that help clinicians determine where a judgment should be examined or verified further. How such deliberation should be presented without imposing an excessive information burden is considered in the following section.

\subsection{Integrating Multi-Agent Reasoning into Clinical Care}
In the clinician evaluation, CHORAL received a significantly higher rating for whether patient evidence supported clinical appraisal, while cross-domain interpretation was also rated higher but did not remain significant after correction. In contrast, no clear differences emerged for the amount and organization of information or for integration into existing clinical workflow. Qualitative findings similarly showed that clinicians found CHORAL's detailed domain reasoning and review--revision process useful for understanding a case, while noting that reviewing the full deliberation during a consultation could impose substantial information burden. Together, these findings suggest that richer reasoning and its effective delivery within clinical workflow are distinct problems. This aligns with prior clinical XAI research showing that the usefulness of explanations depends not only on their availability, but also on how well they fit clinicians' tasks and workflows \cite{tonekaboni2019clinicians,sivaraman2023ignore}. Recent work likewise suggests that richer explanations may improve understanding while increasing decision time or cognitive burden in time-sensitive workflows \cite{hunsicker2026efficiency}.
These findings highlight the need to distinguish \emph{reasoning depth} from \emph{presentation depth} in clinician-facing multi-agent systems. CHORAL's multi-agent process may generate detailed intermediate reasoning, but not all of that reasoning needs to be presented with equal prominence. Our qualitative findings suggest that clinicians were particularly interested in reasoning events that affected how the final judgment should be interpreted, such as consequential disagreements, revised interpretations, and unresolved uncertainty. Routine agreements or repetitive evidence could instead be compressed, while more detailed reasoning could remain available when needed. This type of layered presentation is consistent with prior work on selective transparency in clinical AI, where progressive disclosure has been proposed as a way to provide additional explanation on demand while managing information load \cite{muralidhar2025operationalizing}. Thus, an important design question is not only how much reasoning the system performs, but which parts of that reasoning should be surfaced to clinicians.

When reasoning is presented may be equally important. Participants generally saw greater value in reviewing CHORAL before a visit than in examining the full reasoning process during the encounter itself. Pre-visit review may provide time to examine longitudinal trajectories, domain-level interpretations, disagreements, and revisions in greater depth, whereas in-visit use may require rapid access to key concerns, unresolved issues, and information that still needs clarification. Rather than applying a fixed presentation across all clinical moments, future systems may therefore need to adapt the granularity of reasoning to the workflow stage and the clinician's immediate information needs. This is consistent with prior work on clinical workflow integration, which emphasizes that successful integration depends not merely on placing an AI tool within the workflow, but on providing the right information in an appropriate form and at an appropriate time \cite{salwei2021workflow}.

Finally, better interpretation did not necessarily translate into stronger clinical actionability. Although CHORAL improved support for distinguishing which parts of the system output warranted verification, it did not clearly improve clinicians' ability to identify aspects of the patient case that remained unresolved from the available data or to prioritize prognostic concerns for follow-up. Qualitative findings further suggested that determining follow-up priorities often required clinical context beyond longitudinal PGHD, including the patient's chief complaint, medication and treatment history, recent life events, and ongoing clinical concerns. Participants also noted that suggested follow-up questions often required further adaptation to the individual patient before they could be used directly in a consultation. These findings suggest that interpreting longitudinal PGHD and translating that interpretation into clinical action are related but distinct steps. Prior work in mental-health sensing similarly argues that sensing-derived information must be connected to care context, such as presenting problems, treatment targets, and treatment response, to become clinically actionable \cite{adler2024beyond}. Future systems may therefore need to integrate PGHD-based multi-agent reasoning with broader clinical history and clinician-provided context, rather than relying on PGHD alone to generate a complete follow-up plan.

\subsection{Limitations and Future Work}
This study has several limitations. First, CHORAL's quantitative evaluation used a longitudinal college-student cohort rather than data collected in psychiatric outpatient care, which may limit generalizability to clinical populations and outpatient sensing settings. Second, the evaluation was conducted with a particular LLM configuration and system implementation, and we did not examine whether similar performance and clinician-facing effects would hold across other foundation models or alternative multi-agent architectures. Third, the clinician evaluation compared complete system configurations rather than independently manipulating agent organization, reasoning mechanisms, and presentation. Accordingly, the observed clinician-facing differences should not be interpreted as causal effects of multi-agent organization alone. Fourth, the clinician evaluation was not conducted in an actual clinical setting with clinicians reviewing their own patients; instead, participants assessed researcher-prepared cases. Future work should therefore examine CHORAL in real clinical settings where clinicians can use the system alongside ongoing patient care. Finally, CHORAL primarily relied on longitudinal PGHD and did not incorporate broader clinical context, such as chief complaints, treatment history, medication information, and recent life events, that clinicians identified as important for follow-up decisions. Future work should integrate these additional sources of clinical context and examine whether they improve both prognostic reasoning and support for verification and follow-up prioritization.

\section{Conclusion}
This work explored how multi-agent reasoning can support longitudinal prognostic reasoning from patient-generated health data in mental health follow-up care. We developed CHORAL, a multi-agent scaffold that structures longitudinal trajectory interpretation, domain-specific appraisal, information-gap handling, cross-domain synthesis, and review while keeping the underlying evidence and revisions inspectable. Across a quantitative system evaluation and clinician study, CHORAL showed higher prognostic performance and broader evidence use than a single-agent baseline, while clinicians valued its support for alternative interpretations, contestation, and selective verification. These findings highlight the value of surfacing consequential deliberation while separating reasoning depth from presentation depth in time-constrained clinical workflows.


\bibliographystyle{ACM-Reference-Format}
\bibliography{reference}

\appendix
\section*{Appendix}
\section{Sensor-Derived Sleep State Label Derivation}
\label{appendix-sleep-label}

Since CES does not include a weekly self-reported sleep measure, we derived a binary sleep state from three dimensions over each 7-day window: sleep duration (<6 or >10 hours on average), sleep timing (mean onset before 21:00 or after 02:00), and sleep irregularity (onset- or offset-time
SD >120 minutes). These study-specific operational cutoffs were informed by prior work on sleep duration, sleep-wake timing, and within-person variability~\cite{hirshkowitz2015national,consensus2015recommended, sateia2014international,bei2016beyond,huang2020sleep} and were intended to operationalize pronounced sleep-pattern deviations rather than clinical sleep-disorder thresholds. In CES, sleep onset and offset were encoded in 8-minute units as elapsed time from 20:00, so sleep episodes spanning midnight remained on a continuous scale without clock-time wrap-around. A week was labeled HIGH when at least two of the three dimensions were flagged, and LOW otherwise; this aggregation rule was intended to capture multidimensional sleep-pattern deviations rather than clinical sleep-disorder severity.

\section{CES Feature-to-Behavioral-Marker Mapping}\label{appendix-mapping}
Table~\ref{tab:ces-feature-map} maps the passive sensing features from the CES dataset to higher-level behavioral markers and corresponding WHODAS-informed functioning-relevant domains. To support clinically interpretable reasoning, low-level sensor signals are grouped into symptom-relevant behavioral constructs (e.g., sleep change, concentration, fatigue). These constructs are further aligned with WHODAS-informed functioning-relevant domains based on the WHO Disability Assessment Schedule (WHODAS)~\cite{ustun2010measuring}, enabling a symptom- and function-centered interpretation consistent with clinical practice.
These mappings are informed by prior work in digital phenotyping and behavioral signal interpretation, which associates passive sensing features with clinically relevant constructs such as sleep irregularity, reduced activity, and social withdrawal~\cite{englhardt2024classification, zhang2024reproducible}. This mapping is not intended to imply direct causal relationships, but rather to provide interpretable proxies for behavioral patterns relevant to clinical reasoning. 
The patient agent uses this mapping to organize its output around symptom-functional domains rather than raw sensor streams, enabling downstream Domain Agents to reason over structured behavioral cues. 

\begin{table*}[h]
\centering
\caption{Conceptual mapping of CES passive-sensing features to symptom or behavioral markers and WHODAS-informed functioning-relevant domains. The mapping is used to identify potentially relevant behavioral cues rather than to infer WHODAS scores or establish functional impairment.}
\label{tab:ces-feature-map}
\footnotesize
\renewcommand{\arraystretch}{1.15}
\begin{tabular}{@{} >{\raggedright\arraybackslash}p{0.44\textwidth} >{\raggedright\arraybackslash}p{0.18\textwidth} >{\raggedright\arraybackslash}p{0.32\textwidth} @{}}
\toprule
\textbf{Related Features (from CES)} & \textbf{Symptom / Behavioral Marker} & \textbf{WHODAS-Informed Functioning-Relevant Domain} \\
\midrule

\texttt{sleep\_duration}, \texttt{sleep\_end}, \texttt{sleep\_start}
& Sleep change
& Life activity; Self care \\
\addlinespace

\texttt{loc\_study\_unlock\_duration}, \texttt{loc\_study\_unlock\_num}, \texttt{loc\_home\_unlock\_duration}, \texttt{loc\_home\_unlock\_num}, \texttt{unlock\_num\_hr\_*}, \texttt{other\_playing\_num\_hr\_*}, \texttt{unlock\_num\_ep\_0}, \texttt{other\_playing\_num\_ep\_0}
& Reduced concentration / difficulty concentrating
& Cognition; Life activity \\
\addlinespace

\texttt{loc\_social\_dur}, \texttt{loc\_other\_dorm\_dur}, \texttt{loc\_leisure\_dur}, \texttt{loc\_food\_dur}, \texttt{other\_playing\_duration\_ep\_0}, \texttt{other\_playing\_num\_ep\_0}, \texttt{step\_ep\_0}, \texttt{loc\_dist\_ep\_0}, \texttt{act\_on\_bike\_ep\_0}, \texttt{act\_running\_ep\_0}, \texttt{act\_walking\_ep\_0}, \texttt{loc\_workout\_dur}, \texttt{loc\_home\_dur}, \texttt{loc\_study\_dur}
& Diminished interest / pleasure
& Getting along with others; Participation in society; Self care \\
\addlinespace

\texttt{step\_ep\_0}, \texttt{loc\_dist\_ep\_0}, \texttt{act\_on\_bike\_ep\_0}, \texttt{act\_running\_ep\_0}, \texttt{act\_walking\_ep\_0}, \texttt{step\_hr\_*}, \texttt{loc\_dist\_hr\_*}, \texttt{loc\_workout\_dur}, \texttt{loc\_visit\_num\_ep\_0}
& Fatigue / less energy
& Mobility; Cognition \\
\addlinespace

\texttt{loc\_social\_dur}, \texttt{loc\_other\_dorm\_dur}, \texttt{loc\_visit\_num\_ep\_0}
& Social pattern change
& Getting along with others; Participation in society \\
\addlinespace

\texttt{unlock\_duration\_ep\_0}, \texttt{unlock\_num\_ep\_0}, \texttt{other\_playing\_num\_ep\_0}, \texttt{other\_playing\_duration\_ep\_0}, \texttt{unlock\_duration\_hr\_*}, \texttt{unlock\_num\_hr\_*}, \texttt{other\_playing\_duration\_hr\_*}, \texttt{loc\_home\_unlock\_duration}
& Phone usage pattern change (esp.~nighttime)
& Life activity; Cognition \\

\bottomrule
\end{tabular}
\end{table*}

\section{LLM Implementation and Prompt Templates}
\label{appendix-prompts}

CHORAL was implemented as a multi-agent LLM system using the OpenAI API. All patient, domain, and orchestrator agents used GPT-5-mini as the backbone model. Temperature was not configurable for this model and was therefore not explicitly specified. We set \texttt{\{max\_completion\_tokens\}} to 16,000 per agent invocation, and no completion reached its respective limit. Total computation nevertheless varied across LLM-based conditions, ranging from a mean of 3.6 calls and 77K total tokens per case for the single-agent baseline to 16.1 calls and 392K tokens for CHORAL.

This appendix summarizes the operative prompt templates used in CHORAL and the single-agent baseline. The templates preserve the agent role, task instructions, runtime inputs, relevant background context, reasoning constraints, and structured outputs, while non-operative wording and implementation-specific serialization are omitted for brevity. Runtime variables are denoted by braces (e.g., \texttt{\{patient\_evidence\}}) and instantiated during execution. The prognostic horizon was instantiated as two EMA steps ahead (approximately two weeks) in our experiments. Some prompts were invoked conditionally as part of the reasoning workflow. Complete deployed prompts are provided in the supplementary material.

\subsection{Patient Agent: Trajectory Evidence}
\label{appendix-prompt-patient}

The patient agent constructs a shared, domain-agnostic longitudinal trajectory from the patient's PGHD before domain-specific prognostic appraisal.

\begin{promptbox}{Patient Agent: Trajectory Evidence}

\textbf{Role.}
You are a patient agent that organizes longitudinal behavioral patterns from a phenotype snapshot built from sensor and self-report data. Your role is to construct a shared patient-level trajectory for downstream Domain Agents, rather than to make a domain-specific prognostic judgment.

\textbf{Task.}
Summarize the patient's longitudinal behavioral trajectory by:

1. describing the current behavioral state using statistical, structural, and temporal features;

2. comparing the current state with both the recent and stable baselines;

3. identifying persistence signals indicating that the current behavioral pattern may continue;

4. identifying change signals indicating worsening, recovery, or mixed change;

5. comparing patterns across available observation windows to identify recurrence, persistence, or departure from the patient's prior course; and

6. surfacing functioning-relevant behavioral context when supported by the available evidence.

\textbf{Input.}
Phenotype snapshot with behavioral features over \texttt{\{observation\_windows\}};
\texttt{\{recent\_baseline\}};
\texttt{\{stable\_baseline\}};
longitudinal records from the Phenotype DB; and
\texttt{\{evidence\_catalog\}}.

\textbf{Background / Context.}
Behavioral features may include statistical characteristics, structural changes over time, and temporal patterns across different periods of the day. Evidence identifiers in \texttt{\{evidence\_catalog\}} uniquely identify patient-level evidence and should be preserved for downstream reasoning.

\textbf{Rules.}
Ground each patient-specific statement in the corresponding evidence identifier. When temporal trends and baseline-relative deviations disagree, explicitly preserve the conflict. If evidence for persistence or directional change is weak or conflicting, state this rather than forcing a single interpretation. Do not diagnose a mental-health condition, make a domain-specific prognostic judgment, or infer functional impairment.

\textbf{Output.}
\texttt{behavioral\_summary};
\texttt{baseline\_comparisons};
\texttt{persistence\_signals[]};
\texttt{change\_signals[\{direction\}]};
\texttt{functioning\_relevant\_signals[]};
\texttt{historical\_context[]};
\texttt{cited\_evidence\_ids[]}.

\end{promptbox}

\subsection{Domain Agent: Initial Prognostic Appraisal}
\label{appendix-prompt-domain-initial}

Four Domain Agents corresponding to depression, anxiety, stress, and sleep independently appraise the same shared patient trajectory.

\begin{promptbox}{Domain Agent: Initial Prognostic Appraisal}

\textbf{Role.}
You are a clinical specialist in \texttt{\{domain\}}. Your role is to appraise the patient's likely symptom course for \texttt{\{target\_variable\}} over the \texttt{\{prediction\_horizon\}} from the perspective of your assigned domain.

\textbf{Task.}
Perform a domain-specific prognostic appraisal by:

1. identifying longitudinal evidence relevant to \texttt{\{domain\}}, including trend, recurrence, instability, baseline-relative deviation, and persistence or change;

2. describing functioning-relevant behavioral context when supported by the trajectory;

3. assessing whether the relevant behavioral pattern is likely to persist, worsen, recover, or remain mixed over the prediction horizon;

4. inferring the resulting symptom-course risk and assigning one label from \texttt{\{label\_set\}};

5. using retrieved patient-specific memories, when relevant, to compare the current trajectory with earlier appraisals and their subsequently observed symptom states;

6. identifying evidence that supports or weakens the current appraisal; and

7. identifying unresolved information needs and corresponding follow-up or management considerations.

\textbf{Input.}
Patient Agent trajectory;
\texttt{\{retrieved\_memory\}}, consisting of the \texttt{\{k\_memory\}} most recent confirmed Domain Memory entries when available; and
the evidence identifiers associated with the Patient Agent trajectory.

Each confirmed memory contains the earlier patient trajectory, the previous domain-specific appraisal, the subsequently observed symptom state, and a reflection on how the earlier appraisal and its supporting evidence aligned with or differed from the subsequent course. A citable past-case identifier \texttt{[PC-\ldots]} may be provided for each retrieved memory record.

\textbf{Background / Context.}
\texttt{\{reference\_information\}};
\texttt{\{label\_cutoffs\}} defining the target classes;
\texttt{\{domain\_knowledge\}} containing DSM-informed criteria or the relevant stress framework; and
\texttt{\{ema\_variable\_definitions\}}.

\textbf{Rules.}
This is a prognostic task concerning the symptom course following the observation window rather than a diagnosis or a description of the current state alone. Prioritize longitudinal signals such as trend, recurrence, instability, and baseline-relative change over isolated observations. Consider both supporting and weakening evidence. Interpret the shared trajectory from the perspective of your assigned domain rather than adopting another domain's interpretation. Do not infer functional impairment from sensing evidence alone. Cite patient-specific evidence using its evidence identifier, and cite \texttt{[PC-\ldots]} when a retrieved memory contributes to the appraisal. Treat the appraisal as provisional and commit to one label from \texttt{\{label\_set\}}. Take the current observed state as the default outcome, and depart from it only when at least one escalation or de-escalation signal is cited with its evidence identifier.

\textbf{Output.}
\texttt{predicted\_label};
\texttt{reasoning\{evidence\_trajectory[], functioning\_relevant\_context, prognostic\_outlook, symptom\_risk\_inference\}};
\texttt{supporting\_evidence[]};
\texttt{weakening\_evidence[]};
\texttt{uncertainty\_sources[]};
\texttt{information\_needs[]};
\texttt{follow\_up\_question};
\texttt{intervention\_cue};
\texttt{cited\_evidence\_ids[]}.

\end{promptbox}

\subsection{Domain Memory}
\label{appendix-domain-memory}

After each reasoning round, the Domain Agent stores its domain-specific appraisal as a pending memory entry. Once the corresponding future target state becomes available, the earlier appraisal is compared with the observed course and a reflection is generated describing where the earlier interpretation aligned with or differed from the subsequent outcome, which evidence remained informative, and which aspects of the interpretation were less well supported. The resulting entry becomes a confirmed memory that can be retrieved in subsequent reasoning rounds. In our implementation, each Domain Agent receives the three most recent confirmed memories. A reflection is written when the earlier forecast differed from the observed state; correct forecasts are confirmed with the outcome grade alone. Each Domain Agent additionally receives summary statistics over its own confirmed memories for this patient, covering the outcomes of past departures from the current state and the signals that preceded confirmed transitions. The complete memory construction and reflection prompts are provided in the supplementary material.

\subsection{Information-Gap Handling}
\label{appendix-prompt-information-gap}

When a Domain Agent identifies unresolved information that could clarify its initial appraisal, CHORAL distinguishes information that can be retrieved from the existing longitudinal record from information that requires subsequent clinical follow-up. Internal information needs trigger targeted patient agent retrieval followed by one domain agent refinement, whereas external information needs are retained as clinician follow-up questions. This step is invoked when the initial appraisal departs from the current observed state.

\subsubsection{Domain Agent: Information-Gap Query}
\label{appendix-prompt-domain-query}

\begin{promptbox}{Domain Agent: Information-Gap Query}

\textbf{Role.}
You are a clinical specialist in \texttt{\{domain\}} reviewing unresolved information needs in your current prognostic appraisal.

\textbf{Task.}
Determine what additional information would most help clarify the current appraisal and classify each information need as either internally or externally resolvable.

For an \textbf{internal gap}, formulate a targeted system-level query that can be answered from the existing longitudinal patient record.

For an \textbf{external gap}, formulate a clinician-level follow-up question for information that cannot be obtained from the available PGHD or self-report record.

For each information need, indicate the anticipated course it helps clarify and whether it primarily informs intervention or monitoring.

\textbf{Input.}
Domain Agent appraisal, including \texttt{\{uncertainty\_sources\}} and identified information needs;
Patient Agent report;
available historical context from the longitudinal patient record; and
\texttt{\{allowed\_signals\}} that can be queried from the existing patient data.

\textbf{Background / Context.}
An internal gap may concern whether the current behavioral pattern occurred previously, whether a similar pattern persisted or resolved, or how it previously co-occurred with symptom states. An external gap concerns information not represented in the existing patient data, such as situational context, subjective experience, or the patient's explanation for a behavioral change.

\textbf{Rules.}
Generate exactly \texttt{\{n\_system\}} system follow-ups and \texttt{\{n\_clinician\}} clinician follow-ups when these quantities are specified. System follow-ups must request evidence that can be retrieved from the existing longitudinal record rather than broad resummarization of the full patient history. Clinician follow-ups must concern information that cannot be reliably inferred from the available record. Prefer queries that clarify recurrence, persistence, transition, or recovery when relevant.

Tag each item with the outlook it probes (\texttt{WORSENING}, \texttt{PERSISTENCE}, or \texttt{RECOVERY}) and the planning purpose it informs (\texttt{INTERVENTION} or \texttt{MONITORING}).

\textbf{Output.}
\texttt{system\_followups[\{target\_signal, anticipatory\_focus, planning\_type, purpose\}]};
\texttt{clinician\_followups[\{question, anticipatory\_focus, planning\_type, purpose\}]}.

\end{promptbox}

\subsubsection{Patient Agent: Targeted Evidence Retrieval}
\label{appendix-prompt-patient-retrieval}

For an internally resolvable information need, the patient agent selectively retrieves additional evidence from the patient's longitudinal record rather than reconstructing the full trajectory. Unlike the other steps, this retrieval is implemented as a deterministic procedure over the Phenotype DB rather than as an LLM call, so the returned evidence is restricted to values present in the patient record.

\begin{promptbox}{Patient Agent: Targeted Evidence Retrieval (Procedure)}

\textbf{Input.}
The Domain Agent's system-level query, consisting of a \texttt{\{target\_signal\}} selected from the signals available in the patient record and its stated purpose.

\textbf{Procedure.}
For each system-level query:

1. retrieve the current-window state of the target signal (window statistics, trend, and deviation from the recent and stable baselines);

2. retrieve the same signal from up to five preceding observation windows in the longitudinal record, each assigned a citable past-case identifier \texttt{[PC-...]};

3. determine when the signal departed from the patient's baseline and for how many consecutive windows the deviation has persisted (onset);

4. identify earlier windows with a comparable deviation of the same signal and report how the domain's self-reported symptom state developed in the subsequent window (precedent); and

5. summarize baseline-relative changes in functioning-relevant behavioral signals (functional context).

\textbf{Rules.}
Only information available up to the current reasoning round is used. No domain-specific prognostic judgment is made. If neither current nor historical evidence exists for the requested signal, the query is returned as unresolved.

\textbf{Output.}
\texttt{target\_signal};
\texttt{status} (\texttt{answered} | \texttt{unavailable});
\texttt{current\_signal};
\texttt{prior\_report\_context[\{date, past\_case\_ref, signal\_state, baseline\_deviation\}]};
\texttt{probes\{onset, precedent, functional\_impact\}}.

\end{promptbox}

\subsubsection{Domain Agent: Appraisal Refinement}
\label{appendix-prompt-domain-refinement}

The evidence returned by the patient agent is passed back to the corresponding Domain Agent for a single refinement of its initial appraisal.

\begin{promptbox}{Domain Agent: Appraisal Refinement}

\textbf{Role.}
You are a clinical specialist in \texttt{\{domain\}} refining your provisional prognostic appraisal after receiving additional patient-specific evidence.

\textbf{Task.}
Reconsider the initial appraisal using the retrieved evidence. Determine whether the additional evidence supports retaining or strengthening the initial interpretation, indicates that the interpretation should be revised, or leaves the relevant uncertainty unresolved.

\textbf{Input.}
The original Patient Agent trajectory;
the initial Domain Agent appraisal;
the system-level information need;
the Patient Agent's retrieved evidence; and
\texttt{\{retrieved\_memory\}} when available.

\textbf{Rules.}
Treat the initial appraisal as provisional. Use retrieved evidence only when it is relevant to the identified information need. Consider both evidence that supports and evidence that weakens the initial interpretation. Preserve unresolved uncertainty when the additional evidence is insufficient. Clinician-level follow-up questions are not treated as patient evidence at this stage. Do not introduce patient information that was not provided by the Patient Agent. Preserve relevant citations from the initial appraisal and cite newly retrieved patient evidence using its evidence identifier.

\textbf{Output.}
Same structured output as the initial Domain Agent appraisal, with updated
\texttt{predicted\_label},
\texttt{reasoning},
\texttt{supporting\_evidence[]},
\texttt{weakening\_evidence[]},
\texttt{uncertainty\_sources[]},
\texttt{information\_needs[]}, and
\texttt{cited\_evidence\_ids[]}.

\end{promptbox}

\subsection{Orchestrator: Cross-Domain Synthesis}
\label{appendix-prompt-orchestrator-synthesis}

The Orchestrator combines the four domain-specific appraisals into a provisional integrated assessment while preserving their supporting evidence and differences across domains.

\begin{promptbox}{Orchestrator: Cross-Domain Synthesis}

\textbf{Role.}
You are the Orchestrator coordinating the specialists' domain-specific prognostic appraisals into an integrated assessment over the \texttt{\{prediction\_horizon\}}.

\textbf{Task.}
Construct a provisional cross-domain synthesis by:

1. extracting the major prognostic concerns and supporting evidence from each Domain Agent;

2. comparing agreements and differences across the domain-specific appraisals;

3. identifying behavioral patterns or prognostic concerns supported from multiple domain perspectives;

4. preserving interpretations that remain specific to one domain or differ across domains;

5. examining how evidence or concerns in one domain may contextualize or modulate interpretation in another domain;

6. using retrieved Orchestrator Memory, when relevant, to consider how similar cross-domain patterns previously developed for the same patient;

7. integrating the domain-specific appraisals into a provisional case-level assessment while preserving their underlying evidence and source domain; and

8. generating domain-specific claims from the synthesis for subsequent review by the corresponding Domain Agents.

\textbf{Input.}
\texttt{\{n\_domains\}} Domain Agent appraisals, including their prognostic labels, supporting and weakening evidence, unresolved information needs, and evidence identifiers; and
\texttt{\{retrieved\_memory\}}, consisting of the \texttt{\{k\_memory\}} most recent confirmed Orchestrator Memory entries when available.

Each confirmed Orchestrator Memory contains an earlier final integrated assessment, the subsequently observed symptom states, and a reflection on how the earlier cross-domain interpretation related to the subsequent course.

\textbf{Rules.}
Do not collapse conflicting domain interpretations into consensus merely for consistency. Preserve meaningful differences and unresolved uncertainty. Preserve each domain's own interpretation and its evidence; do not replace one domain's appraisal with another's.
Any modification of a domain-level interpretation must remain grounded in patient-specific evidence and traceable to the original Domain Agent appraisal. Do not invent new patient evidence. Preserve supporting evidence identifiers for substantive patient-specific claims. The Orchestrator records each domain's label, its corroboration, and any revision; a domain label is changed only when a cited escalation or de-escalation signal supports it, and never because other domains are HIGH or LOW.

Each domain claim should identify a specific interpretation that can subsequently be reviewed by the corresponding Domain Agent rather than simply restating its prognostic label.

\textbf{Output.}
\texttt{provisional\_predictions};
\texttt{label\_review};
\texttt{domain\_appraisals};
\texttt{integrated\_case\_formulation};
\texttt{agreements[]};
\texttt{differences[]};
\texttt{cross\_domain\_links[]};
\texttt{domain\_claims};
\texttt{remaining\_uncertainty[]};
\texttt{follow\_up\_needs[]};
\texttt{cited\_evidence\_ids[]}.

\end{promptbox}

\subsection{Domain Agent: Review}
\label{appendix-prompt-domain-review}

Each Domain Agent reviews the portion of the Orchestrator's provisional synthesis relevant to its own domain and identifies interpretations that should be retained, qualified, or revised. For clinician-facing presentation, we grouped the claim-level review verdicts into two categories: \textsc{Agree} for \texttt{CONFIRMED}, and \textsc{Disagree} for \texttt{PARTIAL} or \texttt{RULED\_OUT}. Thus, \textsc{Disagree} indicates that a claim required qualification, revision, or removal rather than necessarily indicating complete rejection.

\begin{promptbox}{Domain Agent: Review}

\textbf{Role.}
You are a clinical specialist in \texttt{\{domain\}} reviewing the Orchestrator's provisional integrated assessment from the perspective of your assigned domain.

\textbf{Task.}
Evaluate each claim under \texttt{\{own\_domain\_claims\}} and determine whether the Orchestrator has accurately represented your original domain-specific appraisal and its supporting evidence.

For each claim, assign one of the following verdicts:

\textbf{CONFIRMED}: The claim is appropriately supported by the available domain-specific evidence.

\textbf{PARTIAL}: The general interpretation is supported, but some part is overstated, insufficiently supported, or requires qualification.

\textbf{RULED\_OUT}: The claim is contradicted by domain-specific evidence or depends on a material misinterpretation of that evidence.

Also identify whether:

1. important supporting or weakening evidence from your original appraisal was omitted;

2. your domain-specific interpretation was oversimplified or merged with another domain in a way that removed an important distinction;

3. relevant uncertainty or follow-up needs were omitted; or

4. evidence from your domain directly conflicts with another domain-specific interpretation in the synthesis.

\textbf{Input.}
Your original Domain Agent appraisal and supporting evidence;
the Orchestrator's provisional integrated assessment; and
\texttt{\{own\_domain\_claims\}}, containing the claims in the synthesis that concern your assigned domain.

\textbf{Rules.}
Evaluate claims from the perspective of your assigned domain. Do not invent new patient evidence. Use existing evidence identifiers when confirming, qualifying, or challenging an interpretation. Preserve domain-specific differences when they are supported by the evidence rather than forcing agreement across agents.

\textbf{Output.}
\texttt{primary\_evaluation\{claims\_evaluated[\{verdict, supporting\_evidence, counter\_evidence\}]\}};
\texttt{omission\_or\_misrepresentation[]};
\texttt{cross\_domain\_notes[\{target\_domain, conflict\_note, severity\}]}.

\end{promptbox}

\subsection{Orchestrator: Revision}
\label{appendix-prompt-orchestrator-revision}

The Orchestrator revises the provisional synthesis using the Domain Agents' review feedback. The synthesis--review--revision process may repeat for up to three rounds.

\begin{promptbox}{Orchestrator: Revision}

\textbf{Role.}
You are the Orchestrator revising an existing integrated assessment after review by the Domain Agents.

\textbf{Task.}
Revise the integrated assessment by:

1. examining the review feedback for each domain-specific claim;

2. preserving interpretations that were confirmed by the corresponding Domain Agent;

3. refining or qualifying claims receiving a \texttt{PARTIAL} verdict;

4. removing or revising claims receiving a \texttt{RULED\_OUT} verdict when they are inconsistent with the corresponding domain evidence;

5. restoring important domain-specific findings, uncertainties, or information needs that were omitted or misrepresented;

6. incorporating cross-domain conflicts when they materially affect the integrated case formulation; and

7. updating domain-level or integrated conclusions when warranted by the review feedback and supporting evidence.

\textbf{Input.}
Current integrated assessment and its \texttt{\{domain\_claims\}};
the original Domain Agent appraisals and supporting evidence;
the Domain Agents' claim-level evaluations;
\texttt{\{omission\_or\_misrepresentation\}} feedback; and
\texttt{\{cross\_domain\_notes\}}.

\textbf{Background / Context.}
The revised synthesis may be returned to the Domain Agents for another review. The synthesis--review--revision process repeats for up to \texttt{\{max\_rounds\}} rounds; \texttt{\{max\_rounds\}} was set to three in our implementation.

\textbf{Rules.}
Do not introduce new patient evidence. Revisions must be grounded in the original Domain Agent appraisals, patient evidence, and Domain Agent review feedback. Preserve confirmed claims unless additional review feedback provides evidence requiring reconsideration. Preserve meaningful disagreement across domains when the available evidence does not support reconciliation. Label changes follow the same evidence conditions as the synthesis stage. The process may terminate before \texttt{\{max\_rounds\}} when no further substantive revision is required.

\textbf{Output.}
Updated
\texttt{provisional\_predictions};
\texttt{domain\_appraisals};
\texttt{integrated\_case\_formulation};
\texttt{agreements[]};
\texttt{differences[]};
\texttt{cross\_domain\_links[]};
\texttt{domain\_claims};
\texttt{remaining\_uncertainty[]};
\texttt{follow\_up\_needs[]}; and
\texttt{cited\_evidence\_ids[]}.

\end{promptbox}

\subsection{Orchestrator Memory}
\label{appendix-orchestrator-memory}

After the review--revision process is completed, the final integrated assessment is stored as a pending Orchestrator Memory entry. Once the corresponding future target states become available, the earlier integrated assessment is compared with the observed cross-domain course and a reflection is generated. The resulting confirmed memory preserves the earlier integrated interpretation, subsequent target states, and outcome reflection for use in later synthesis. In our implementation, the Orchestrator receives the three most recent confirmed memories. The complete memory construction and reflection prompts are provided in the supplementary material.

\subsection{Clinician-Facing Final Report}
\label{appendix-prompt-final-report}

After the review--revision process, the Orchestrator converts the completed integrated assessment into the structured report presented through the Clinical Dashboard. The report prioritizes the patient's recent longitudinal trajectory and final domain-level prognostic appraisals, followed by their supporting evidence, important cross-domain agreements or differences, remaining uncertainty, follow-up questions, and intervention or management considerations. The reporting stage is instructed not to introduce new patient evidence or independently alter the completed integrated assessment. The complete reporting prompt is provided in the supplementary material.

\subsection{Single Agent Baseline}
\label{appendix-prompt-single-agent}

The single-agent baseline retained the same patient agent trajectory, domain-specific knowledge, target definitions, output requirements, and comparable longitudinal memory used by CHORAL. It removed CHORAL's decomposition into separate domain agents, its on-demand patient agent retrieval and domain agent refinement process, and its orchestrator-mediated synthesis, review, and revision. The patient agent trajectory stage was retained and used the same template as in CHORAL. A single GPT-5-mini instance then performed the four domain-specific appraisals and cross-domain integration within a single reasoning pass.

\begin{promptbox}{Single Agent Baseline}

\textbf{Role.}
You are a single clinical reasoning agent forecasting the patient's symptom course across all \texttt{\{n\_domains\}} domains over the \texttt{\{prediction\_horizon\}}. Perform the domain-specific appraisals and cross-domain integration yourself without dividing the task among separate specialist agents.

\textbf{Task.}

1. \textbf{Per-domain appraisal.}
Evaluate each domain separately from the shared patient trajectory. For each domain, identify longitudinal supporting and weakening evidence, functioning-relevant context, the expected symptom course over the prediction horizon, relevant patient-specific history, unresolved information needs, and follow-up or intervention considerations. Give each domain comparable reasoning depth.

2. \textbf{Cross-domain synthesis.}
Integrate the domain-specific appraisals by examining shared evidence, agreements, differences, and relevant relationships across domains. Preserve meaningful differences when the available evidence does not support a unified interpretation.

3. \textbf{Finalization.}
Produce one prognostic label for each domain and an integrated case-level assessment.

\textbf{Input.}
\texttt{\{patient\_evidence\}}, consisting of the same Patient Agent trajectory used by CHORAL, including baseline comparisons, longitudinal trajectory summaries, functioning-relevant behavioral context, and evidence identifiers;
\texttt{\{evidence\_catalog\}}; and
\texttt{\{retrieved\_memory\}}, consisting of the \texttt{\{k\_memory\}} most recent confirmed prediction--outcome records from a unified memory store.

Each retrieved memory contains earlier domain-specific prognostic appraisals, subsequently observed symptom states, and reflections on how the earlier appraisals and supporting evidence related to the subsequent course. A citable past-case identifier \texttt{[PC-\ldots]} may be provided when a retrieved record contributes to the current reasoning.

\textbf{Background / Context.}
\texttt{\{domain\_knowledge\}} for all \texttt{\{n\_domains\}} domains;
\texttt{\{label\_cutoffs\}} for each domain;
\texttt{\{ema\_variable\_definitions\}}; and
\texttt{\{label\_set\}}.

The Patient Agent trajectory is generated in a separate preceding call using the same Patient Agent template as CHORAL.

\textbf{Rules.}
This is a prognostic task rather than a description of the current state alone. Prioritize longitudinal signals such as trend, recurrence, instability, and baseline-relative change over isolated observations. Consider both supporting and weakening evidence for each domain. Evaluate each domain from its own perspective rather than carrying one domain's conclusion over to another. Take the current observed state as the default outcome for each domain, and depart from it only when at least one escalation or de-escalation signal is cited with its evidence identifier.

Ground patient-specific interpretations with their evidence identifiers and do not invent identifiers. When retrieved memory contributes to the reasoning, interpret the earlier appraisal together with its subsequently observed symptom state and reflection and cite the corresponding \texttt{[PC-\ldots]} identifier. Commit to one label from \texttt{\{label\_set\}} for each domain.

Identify unresolved information needs from the available evidence, but do not initiate additional Patient Agent retrieval or iterative appraisal refinement.

Do not introduce separate specialist personas, an Orchestrator, inter-agent discussion, specialist review, or iterative revision.

\textbf{Output.}
\texttt{recent\_behavioral\_summary};
\texttt{domain\_assessments\{label, reasoning\{evidence\_trajectory[], functioning\_relevant\_context, prognostic\_outlook, symptom\_risk\_inference\}, supporting\_evidence[], weakening\_evidence[], uncertainty\_sources[], information\_needs[], follow\_up\_question, intervention\_cue\}};
\texttt{cross\_domain\_links[]};
\texttt{integrated\_synthesis};
\texttt{information\_gaps[]};
\texttt{followup\_questions[]};
\texttt{intervention\_cues[]};
\texttt{cited\_evidence\_ids[]};
\texttt{final\_predictions}.

\end{promptbox}

\clearpage
\section{Clinician Participant Demographics}
\label{appendix-participants}

\begin{table*}[!ht]
\centering
\small
\caption{Demographic characteristics of clinician participants ($N=22$). Clinical experience includes internship, residency, and subsequent clinical practice.}
\label{tab:clinician-demographics}

\begin{tabular}{cccll}
\toprule
\textbf{Participant} &
\textbf{Age} &
\textbf{Gender} &
\textbf{Professional Status} &
\textbf{Clinical Experience} \\
\midrule

P1  & 46 & M & Board-certified psychiatrist & $>$10 years \\
P2  & 48 & M & Board-certified psychiatrist & $>$10 years \\
P3  & 37 & M & Board-certified psychiatrist & $>$10 years \\
P4  & 44 & M & Board-certified psychiatrist & $>$10 years \\
P5  & 37 & F & Board-certified psychiatrist & 10 years \\
P6  & 49 & M & Board-certified psychiatrist & $>$10 years \\

P7  & 34 & F & Psychiatry resident & 5 years \\
P8  & 31 & M & Psychiatry resident & 5 years \\
P9  & 31 & F & Psychiatry resident & 5 years \\
P10 & 33 & M & Board-certified psychiatrist & 7 years \\
P11 & 33 & M & Psychiatry resident & 3 years \\
P12 & 33 & M & Psychiatry resident & 3 years \\
P13 & 33 & M & Board-certified psychiatrist & 7 years \\
P14 & 28 & F & Psychiatry resident & 3 years \\
P15 & 28 & F & Psychiatry resident & 2 years \\
P16 & 30 & M & Psychiatry resident & 4 years \\
P17 & 31 & M & Psychiatry resident & 4 years \\
P18 & 32 & M & Psychiatry resident & 4 years \\
P19 & 30 & F & Psychiatry resident & 3 years \\
P20 & 29 & F & Psychiatry resident & 3 years \\
P21 & 33 & M & Psychiatry resident & 2 years \\
P22 & 30 & M & Psychiatry resident & 3 years \\

\bottomrule
\end{tabular}
\end{table*}
\FloatBarrier

\section{Clinician Questionnaire}
\label{appendix-questionnaire}

Table~\ref{tab:clinician-questionnaire} lists the questionnaire items used to assess clinical usefulness, actionability, workflow fit, contestability and reliance support. All items were rated on a 7-point Likert scale.

\begin{table*}[!ht]
\centering
\small
\caption{Questionnaire items used in the clinician evaluation.}
\label{tab:clinician-questionnaire}

\begin{tabular}{
>{\raggedright\arraybackslash}p{0.22\textwidth}
>{\raggedright\arraybackslash}p{0.72\textwidth}
}
\toprule
\textbf{Measure} & \textbf{Questionnaire Item} \\
\midrule

\multicolumn{2}{l}{\textit{\textbf{RQ2: Clinical Usefulness, Clinical Actionability, and Workflow Fit}}} \\
\midrule

\multirow{3}{*}{Clinical Usefulness}
&
1. The prognostic judgment was clinically useful given the patient's recent changes.
\\
&
2. The supporting evidence helped me clinically appraise the validity of the prognostic judgment.
\\
&
3. Integrating findings across multiple symptom domains was useful for interpreting the patient's overall state.
\\

\midrule

\multirow{2}{*}{Clinical Actionability}
&
4. The presented information helped me identify aspects that were difficult to determine from the available data and required further verification.
\\
&
5. The presented information helped me determine which prognostic concerns should be prioritized and why.
\\

\midrule

\multirow{2}{*}{Workflow Fit}
&
6. The amount and organization of information were appropriate for review within actual clinical practice.
\\
&
7. The system output could be integrated into existing clinical workflow as a reference for prognostic judgment and follow-up planning.
\\

\midrule

\multicolumn{2}{l}{\textit{\textbf{RQ3: Contestability and Reliance Support}}} \\
\midrule

\multirow{3}{*}{Contestability Support}
&
8. I could trace which patient data and domain-specific findings supported the final prognostic judgment.
\\
&
9. I could identify points where findings across domains disagreed or where alternative interpretations were possible.
\\
&
10. If I disagreed with the system's judgment, I could identify which evidence or domain-specific judgments should be further reviewed or challenged.
\\

\midrule

\multirow{3}{*}{\shortstack[l]{Reliance and\\ Verification Support}}
&
11. I felt that the system's prognostic judgment could be relied on as a clinical reference.
\\
&
12. I could distinguish which parts of the system's judgment I could rely on and which required further verification.
\\
&
13. I felt that the information provided by the system was unlikely to lead to clinically inappropriate or unsafe judgments.
\\

\bottomrule
\end{tabular}
\end{table*}
\FloatBarrier

\section{Single Agent Dashboard}
\begin{figure*}[t]
    \centering
    \includegraphics[width=\textwidth]{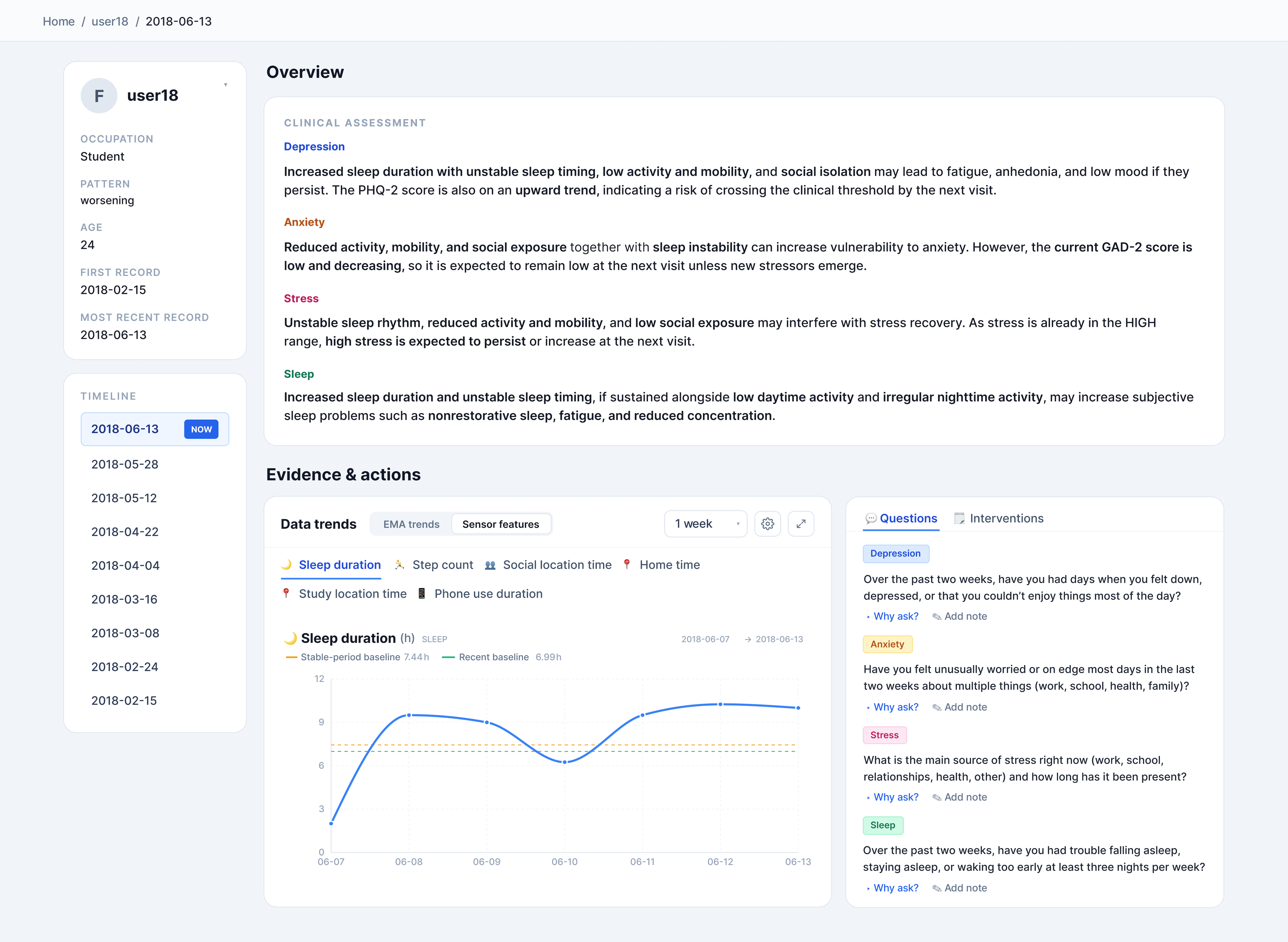}
    \caption{Dashboard used in the single-agent condition during the clinician evaluation.}
    \label{fig:baseline-dashboard}
\end{figure*}

The single-agent condition used a dashboard with a similar high-level organization to the CHORAL interface, while omitting the intermediate reasoning views that were specific to CHORAL's multi-agent process. As described in Section~\ref{main-study-design}, the single-agent generated the four domain appraisals and integrated assessment within a single LLM generation, which were presented together with the corresponding supporting evidence and follow-up information. Figure~\ref{fig:baseline-dashboard} shows the dashboard used in the single-agent condition. 

\section{Domain Performance}

\subsection{Domain-level Prognostic Performance}
\label{appendix-domain-performance}

\begin{table*}[t]
\centering
\small
\caption{Domain-level prognostic performance for depression.}
\label{tab:domain-depression}
\begin{tabular}{lcccccc}
\toprule
\textbf{Condition} &
\textbf{BalAcc} &
\textbf{Precision} &
\textbf{Recall} &
\textbf{F1} &
\textbf{Accuracy} &
\textbf{Deterioration Detection} \\
\midrule
Majority Baseline
& 0.500 & 0.000 & 0.000 & 0.000 & \textbf{0.904} & 0/42 (0\%) \\
Persistence Baseline
& 0.676 & \textbf{0.433} & 0.408 & 0.420 & 0.892 & 0/42 (0\%) \\
Single Agent Baseline
& 0.689 & 0.295 & 0.507 & 0.373 & 0.836 & 11/42 (26\%) \\
w/o Orchestrator
& 0.701 & 0.285 & \textbf{0.549} & 0.375 & 0.824 & \textbf{12/42 (29\%)} \\
w/o Memory
& 0.681 & 0.352 & 0.451 & 0.395 & 0.868 & 4/42 (10\%) \\
w/o On-demand Retrieval
& 0.672 & 0.311 & 0.451 & 0.368 & 0.851 & 8/42 (19\%) \\
CHORAL
& \textbf{0.711} & 0.387 & 0.507 & \textbf{0.439} & 0.876 & 9/42 (21\%) \\
\bottomrule
\end{tabular}
\end{table*}

\begin{table*}[t]
\centering
\small
\caption{Domain-level prognostic performance for anxiety.}
\label{tab:domain-anxiety}
\begin{tabular}{lcccccc}
\toprule
\textbf{Condition} &
\textbf{BalAcc} &
\textbf{Precision} &
\textbf{Recall} &
\textbf{F1} &
\textbf{Accuracy} &
\textbf{Deterioration Detection} \\
\midrule
Majority Baseline
& 0.500 & 0.000 & 0.000 & 0.000 & \textbf{0.880} & 0/48 (0\%) \\
Persistence Baseline
& 0.693 & \textbf{0.456} & 0.461 & \textbf{0.458} & 0.869 & 0/48 (0\%) \\
Single Agent Baseline
& 0.667 & 0.295 & 0.494 & 0.370 & 0.797 & 7/48 (15\%) \\
w/o Orchestrator
& 0.682 & 0.315 & 0.517 & 0.391 & 0.807 & \textbf{12/48 (25\%)} \\
w/o Memory
& 0.680 & 0.408 & 0.449 & 0.428 & 0.855 & 1/48 (2\%) \\
w/o On-demand Retrieval
& 0.678 & 0.296 & \textbf{0.528} & 0.379 & 0.792 & \textbf{12/48 (25\%)} \\
CHORAL
& \textbf{0.695} & 0.359 & 0.517 & 0.424 & 0.831 & 10/48 (21\%) \\
\bottomrule
\end{tabular}
\end{table*}

\begin{table*}[t]
\centering
\small
\caption{Domain-level prognostic performance for stress.}
\label{tab:domain-stress}
\begin{tabular}{lcccccc}
\toprule
\textbf{Condition} &
\textbf{BalAcc} &
\textbf{Precision} &
\textbf{Recall} &
\textbf{F1} &
\textbf{Accuracy} &
\textbf{Deterioration Detection} \\
\midrule
Majority Baseline
& 0.500 & 0.000 & 0.000 & 0.000 & \textbf{0.792} & 0/92 (0\%) \\
Persistence Baseline
& 0.621 & \textbf{0.397} & 0.403 & 0.400 & 0.749 & 0/92 (0\%) \\
Single Agent Baseline
& 0.616 & 0.324 & 0.513 & 0.397 & 0.676 & 26/92 (28\%) \\
w/o Orchestrator
& 0.647 & 0.323 & \textbf{0.656} & 0.433 & 0.642 & \textbf{40/92 (43\%)} \\
w/o Memory
& 0.624 & 0.339 & 0.506 & 0.406 & 0.692 & 18/92 (20\%) \\
w/o On-demand Retrieval
& 0.606 & 0.306 & 0.526 & 0.387 & 0.653 & 31/92 (34\%) \\
CHORAL
& \textbf{0.651} & 0.341 & 0.617 & \textbf{0.439} & 0.672 & 34/92 (37\%) \\
\bottomrule
\end{tabular}
\end{table*}

\begin{table*}[t]
\centering
\small
\caption{Domain-level prognostic performance for sleep.}
\label{tab:domain-sleep}
\begin{tabular}{lcccccc}
\toprule
\textbf{Condition} &
\textbf{BalAcc} &
\textbf{Precision} &
\textbf{Recall} &
\textbf{F1} &
\textbf{Accuracy} &
\textbf{Deterioration Detection} \\
\midrule
Majority Baseline
& 0.500 & 0.000 & 0.000 & 0.000 & 0.576 & 0/113 (0\%) \\
Persistence Baseline
& \textbf{0.698} & \textbf{0.659} & 0.640 & \textbf{0.649} & \textbf{0.707} & 0/113 (0\%) \\
Single Agent Baseline
& 0.584 & 0.490 & 0.717 & 0.582 & 0.564 & 60/113 (53\%) \\
w/o Orchestrator
& 0.598 & 0.491 & 0.828 & 0.617 & 0.564 & 73/113 (65\%) \\
w/o Memory
& 0.578 & 0.473 & \textbf{0.879} & 0.615 & 0.532 & \textbf{80/113 (71\%)} \\
w/o On-demand Retrieval
& 0.603 & 0.505 & 0.736 & 0.599 & 0.582 & 78/113 (69\%) \\
CHORAL
& 0.633 & 0.524 & 0.806 & 0.635 & 0.607 & 65/113 (58\%) \\
\bottomrule
\end{tabular}
\end{table*}

Tables~\ref{tab:domain-depression}--\ref{tab:domain-sleep} report balanced accuracy, precision, recall, F1, accuracy, and deterioration detection for each of the four target domains. Deterioration refers to a transition from a LOW to a HIGH state. Overall, CHORAL achieved the highest balanced accuracy for depression (0.711), anxiety (0.695), and stress (0.651), as well as the highest F1 for depression and stress (0.439). Across these three domains, the Majority baseline remained at chance-level balanced accuracy (0.500) and detected no deterioration, while the Persistence baseline also detected no deterioration because carrying forward the current label cannot anticipate a LOW-to-HIGH transition. In contrast, CHORAL detected deterioration cases across all three domains. Sleep was the only domain in which the Persistence baseline achieved higher balanced accuracy than CHORAL, as discussed in the main text.

The deterioration detection results further illustrate this difference. All LLM-based conditions detected deterioration cases across the four domains, whereas the Majority and Persistence baselines detected none. CHORAL detected 21\% of deterioration cases in depression, 21\% in anxiety, 37\% in stress, and 58\% in sleep. The ablation results also show different recall--specificity trade-offs across configurations. In particular, removing the orchestrator generally increased recall while reducing specificity, whereas full CHORAL achieved the highest balanced accuracy in three of the four domains.

\section{Run-to-Run Variability Analysis}
To assess run-to-run variability, we independently re-executed CHORAL once under the same settings and compared prognostic performance, evidence use, domain-level predictions, and referenced evidence identifiers. Evidence-ID agreement was computed over binary referenced/not-referenced codes for each identifier in each case-week's evidence catalog ($N=34{,}188$), pooling references across the four domain-level reasonings in the Orchestrator's final output.

The two runs yielded similar balanced accuracy (0.673 vs. 0.672) and evidence channel coverage (89.2\% vs. 89.0\%). Domain-level predictions showed 88.7\% agreement ($\kappa=0.745$), and referenced evidence identifiers showed 90.8\% agreement ($\kappa=0.812$). Jointly, aggregate performance was similar across the two runs.

\end{document}